\documentclass[paper,notoc]{JHEP3}

\usepackage{amsmath}
\usepackage{amsfonts}
\usepackage{amssymb}
\usepackage{graphics}
\usepackage{epsfig}
\usepackage{axodraw4j}

\def\beq{\begin{equation}}
\def\eeq{\end{equation}}
\newcommand{\bea}{\begin{eqnarray}}
\newcommand{\eea}{\end{eqnarray}}

\def\bsp#1\esp{\begin{split}#1\end{split}}

\newcommand{\eps}{\epsilon}

\newcommand{\ord}{\begin{cal}O\end{cal}}
\newcommand{\cI}{\begin{cal}I\end{cal}}

\newcommand{\cB}{{\cal B}}

\newcommand{\cS}{{\cal S}}

\newcommand{\cP}{{\cal P}}
\newcommand{\cQ}{{\cal Q}}

\def\cQ{{\cal Q} }
\def\cP{{\cal P} }

\def\cN{{\cal N} }

\def\cZ{{\cal Z} }
\def\cI{{\cal I} }

\newcommand{\rd}{\mathrm{d}}

\newcommand{\cH}{\begin{cal}H\end{cal}}

\renewcommand{\log}{\ln}

\def\bit#1\eit{\begin{itemize}#1\end{itemize}}
\def\ben#1\een{\begin{enumerate}#1\end{enumerate}}

\newenvironment{sloppyequation}[0]{\sloppy\begin{flushleft}\hspace*{0.75cm}\(}{\)\end{flushleft}\fussy}
\newenvironment{sloppytext}[0]{\sloppy\begin{flushleft}}{\end{flushleft}\fussy}

\newcommand{\beqsloppy}{\begin{sloppyequation}}
\newcommand{\eeqsloppy}{\end{sloppyequation}}
\newcommand{\btxtsloppy}{\begin{sloppytext}}
\newcommand{\etxtsloppy}{\end{sloppytext}}

\newcommand{\li}{\textrm{Li}}

\newcommand{\zp}{\bar{z}}

\title{Three-mass triangle integrals and single-valued polylogarithms}

\author{Federico Chavez, Claude Duhr\\
Institut f\"ur theoretische Physik, ETH Z\"urich,\\ Wolfgang-Paulistr. 27, CH-8093, Switzerland\\
E-mail:~\email{fchavez@itp.phys.ethz.ch, duhrc@itp.phys.ethz.ch}}

\abstract{We study one and two-loop triangle integrals with massless propagators and all external legs off shell. 
We show that there is a kinematic region where the results can be expressed in terms of a basis of single-valued polylogarithms
in one complex variable. The relevant space of single-valued functions can be determined a priori and the results take 
strikingly a simple and compact form when written in terms of this basis. We study the properties of the basis functions 
and illustrate how one can easily analytically continue our results to all kinematic regions where the external masses have the same sign.}


\begin{document}


\section{Introduction}
\label{sec:intro}
The evaluation of Feynman integrals is a necessary ingredient to higher-order corrections to physical observables in quantum field theories.
For this reason analytic computations of two-loop Feynman integrals have seen a lot of interest over the last decade. The general strategy currently consists in reducing all 
the Feynman diagrams that contribute to a given process to small set of so-called master integrals by using integration-by-parts and Lorentz invariance identities~\cite{Chetyrkin:1980pr,Chetyrkin:1981qh,Gehrmann:1999as,Laporta:2001dd}. Various techniques have been developed over the last decade for the computation of the master integrals, the most prominent ones probably being the differential equation~\cite{Kotikov:1990kg,Kotikov:1991hm,Kotikov:1991pm} and Mellin-Barnes approaches~\cite{Boos:1990rg}. All of these procedures finally lead to analytical results for the master integral in the form of a Laurent series in the dimensional regulator $\eps=(4-D)/2$ whose coefficients are transcendental functions of the scales of the process. While there are examples of multi-loop integrals which evaluate to elliptic integrals~\cite{Laporta:2004rb}, it is known that for large classes of Feynman integrals these transcendental functions consist only in polylogarithms and generalizations thereof~\cite{Remiddi:1999ew,Gehrmann:2000zt,Goncharov:1998, Goncharov:2001,Davydychev:2003mv,Davydychev:2000na,Ablinger:2011te}.

Focusing on two-loop $2\to2$ processes relevant to hadron collider physics, all the two-loop master integrals with massless propagators for the production of two massless particles~\cite{Smirnov:1999gc,Tausk:1999vh,Smirnov:1999wz,Anastasiou:2000mf,Anastasiou:2000kp} or one massive and one massless particle~\cite{Gehrmann:2000zt,Gehrmann:2001ck} have been computed analytically. Recently, master integrals for the production of a pair of heavy quarks at hadron colliders have also become available~\cite{Bonciani:2008az,Bonciani:2009nb,Bonciani:2010mn}. 

The focus of this paper are two-loop three-mass triangle integrals without internal masses. These integrals are for example relevant for the computation of the two-loop corrections to the production of two heavy particles with different masses. A prominent example of such a process is the production of a pair of weak gauge bosons at a hadron collider, an important background to many BSM and Higgs searches at the Large Hadron Collider (LHC). In ref.~\cite{Birthwright:2004kk} it was shown that all two-loop three-mass triangle integrals with massless propagators can be reduced to a small set of master integrals consisting of one and two-loop two and three-point functions. In this paper we present analytical expressions for all the three-point master integrals presented in ref.~\cite{Birthwright:2004kk}. While this is not the first time that these integrals have been evaluated analytically, we believe that our results go beyond existing representations available in the literature. Indeed, while some of the master integrals were already evaluated almost two decades ago in refs.~\cite{Usyukina:1992jd,Usyukina:1994iw,Davydychev:1999mq} in terms of classical polylogarithms, not all of them were computed to the order in the $\eps$ expansion required for two-loop computations in general, i.e., up to terms in the $\eps$ expansion of transcendental weight four. In ref.~\cite{Birthwright:2004kk} the two-loop master integrals were computed up to weight four using the differential equations approach and the results were expressed in terms of complicated iterated integrals whose integration kernels involve inverse square roots of a K\"allen function. The resulting set of functions is then not related to polylogarithms in a straightforward way.

The aim of this paper is to present fully analytic results for all two-loop three-mass triangle integrals without internal masses in terms of multiple polylogarithms up to weight four. We show that, using generic considerations on the analytic structure of these integrals, one can identify a priori a basis for a space of polylogarithmic functions through which all of these integrals can be expressed. This space of functions is composed of single-valued functions of a single complex variable $z$ (and its complex conjugate $\zp$), and encompasses in particular the famous Bloch-Wigner dilogarithm and the single-valued versions of the harmonic polylogarithms introduced in ref.~\cite{BrownSVHPLs}. The latter functions were recently used to derive analytic results for the six-point amplitude in $\cN=4$ Super Yang-Mills in the multi-Regge limit~\cite{Dixon:2012yy} and for certain generalized ladder integrals~\cite{Drummond:2012bg} for a high number of loops. In terms of this basis of functions, all our results are characterized by very compact analytic expressions that make all the symmetries and analytic continuations manifest. In particular, we observe that one of the two-loop master integrals can be expressed, at least up to transcendental weight four, as a combination of the one-loop triangle and the two-loop ladder integral. This extends an observation made in ref.~\cite{Usyukina:1994iw} to one order higher in the $\eps$ expansion.

This paper is organized as follows: In section~\ref{sec:kinematics} we define our notations and conventions and discuss some general properties of three-mass triangle integrals. In particular we argue that, in a specific kinematic region, they can naturally be expressed through a certain class of single-valued functions of a single complex variable $z$ and its complex conjugate $\zp$, and in section~\ref{sec:SV_polylogs} we give a short review of these functions up to weight four. In section~\ref{sec:generic_z} we present our results in the kinematic region where the functions are single-valued, and in section~\ref{sec:anal_cont} we perform the analytic continuation of our results to other kinematic regions.
We include appendices that contain details about the basis of single-valued functions.


\section{Triangle integrals with three external masses}
\label{sec:kinematics}
We start by discussing the kinematics of three-point functions where all three external legs are off shell and all internal propagator are massless. If $T^{(\ell)}(p_1,p_2,p_3;\epsilon)$ denotes a generic $\ell$-loop integral of this type in $D=4-2\epsilon$ dimensions with external momenta $p_i$, $i=1,2,3$, then Lorentz invariance and momentum conservation imply that the result can only depend on the virtualities $p_i^2\neq 0$. In dimensional regularization we can therefore write, without loss of generality,
\beq\label{eq:generic_triangle}
T^{(\ell)}(p_1,p_2,p_3;\eps) = c_{\Gamma}^\ell\,(-p_3^2)^{n-\ell \eps}\,\mathcal{T}^{(\ell)}(u,v;\eps)\,,
\eeq
for some integer $n$ and where we defined
\beq\label{eq:u_v_def}
u = {p_1^2\over p_3^2} {\rm~~and~~} v = {p_2^2\over p_3^2} \,.
\eeq
In eq.~\eqref{eq:generic_triangle} we pulled out the usual loop factor
\beq
c_\Gamma=e^{\gamma_E\eps}{\Gamma(1+\eps)\,\Gamma(1-\eps)^2\over\Gamma(1-2\eps)}\,,
\eeq
where $\gamma_E = -\Gamma'(1)$ is the Euler-Mascheroni constant. In the following, and unless stated otherwise, we will always work in the Euclidean region where $p_i^2<0$, $i=1,2,3$, and thus $u,v>0$. All the results we present in this paper are real in the Euclidean region. We note that our results will be equally valid in the physical region $p_i^2>0$. This region is phenomenologically relevant for two reasons. Firstly, it describes the decay of a heavy particle of mass $p_3^2$ into two lighter particles\footnote{Without loss of generality we may assume that $p_3^2$ is the largest invariant.}. Secondly, three-mass triangle integrals in this region appear in the production amplitude for a pair of weak gauge bosons at higher orders in perturbation theory.  The region $p_i^2>0$ is related to the Euclidean region via the analytic continuation
\beq\label{eq:anal_cont}
-(p_k^2+i\varepsilon)  \to  e^{-i\pi} \left|p_k^2\right|\,.
\eeq
It is then easy to see that the phase factors cancel out in the ratios~\eqref{eq:u_v_def} so that the analytic continuation of eq.~\eqref{eq:generic_triangle} from the Euclidean to the physical region $p_i^2>0$ is trivial,
\beq
T^{(\ell)}(p_1,p_2,p_3;\eps) \to (-1)^n\,e^{-i\pi\ell\eps}\,c_{\Gamma}^\ell\,\left|p_3^2\right|^{n-\ell \eps}\,\mathcal{T}^{(\ell)}(u,v;\eps)\,.
\eeq

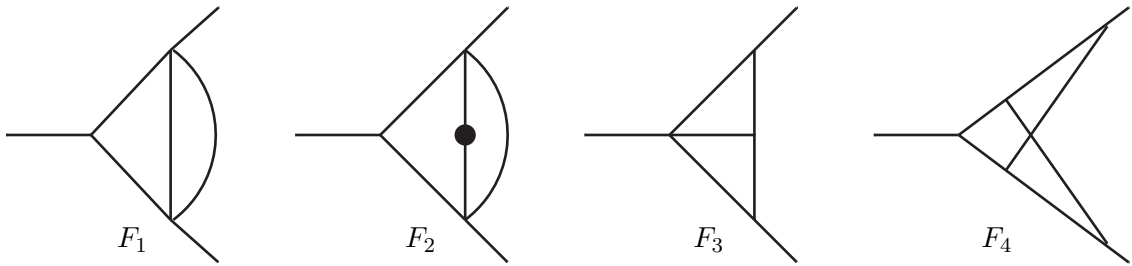
\begin{figure}[!t]
\begin{center}
  \begin{picture}(472,98) (10,-15)	   
    \SetWidth{1.0}
    \Line(10,34)(42,34)
    \Line(42,34)(72,66)
    \Line(72,66)(72,2)
    \Line(72,2)(42,34)
    \Line(90,82)(72,66)
    \Line(72,2)(90,-14)
    \Arc[clock](49,34)(40,53.13,-53.13)
    \Line(119,34)(151,34)
    \Line(151,34)(183,66)
    \Line(183,66)(183,2)
    \Line(183,2)(151,34)
    \Line(199,82)(183,66)
    \Line(183,2)(199,-14)
    \Arc[clock](159,34)(40,53.13,-53.13)
    \Vertex(183,34){4}
    \Text(52,-10)[lb]{\normalsize{{$F_1$}}}
    \Text(161,-10)[lb]{\normalsize{{$F_2$}}}
    \Text(270,-10)[lb]{\normalsize{{$F_3$}}}
    \Text(379,-10)[lb]{\normalsize{{$F_4$}}}
    \Line(337,34)(369,34)
    \Line(433,-14)(369,34)
    \Line(433,82)(369,34)
    \Line(387,47)(425,-7)
    \Line(387,21)(425,75)
    \Line(228,34)(260,34)
    \Line(260,34)(292,66)
    \Line(292,66)(292,2)
    \Line(292,2)(260,34)
    \Line(308,82)(292,66)
    \Line(292,2)(308,-14)
    \Line(292,34)(260,34)
  \end{picture}	
\end{center}
\caption{\label{fig:master_integrals}The master integrals for the two-loop three-mass triangle integrals.}
\end{figure}

The main focus of this paper are three-mass triangles that appear in two-loop computations in dimensional regularization. In ref.~\cite{Birthwright:2004kk} it was shown that all two-loop three-mass triangles can be reduced to a limited set of master integrals (see fig.~\ref{fig:master_integrals}). The 
reduction to master integrals involves also two-loop two-point functions which can be evaluated to all orders in $\eps$ in terms of $\Gamma$ functions. We will therefore not consider the two-point master integrals any further and will concentrate exclusively on the three-point functions.

\begin{figure}[!t]
\begin{center}
\includegraphics[scale=0.44]{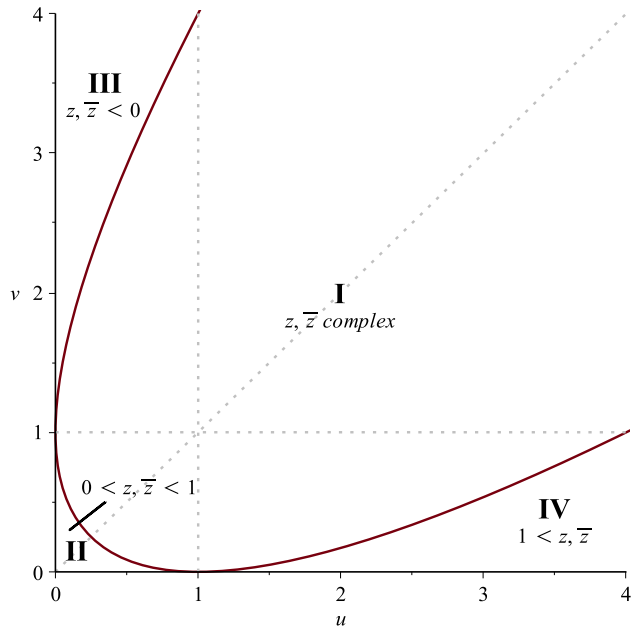}
\caption{\label{fig:regions}Different regions in $(u,v)$ space. The parabola represents the curve $\lambda(1,u,v)=0$.}
\end{center}
\end{figure}

As we will see in the next sections, the kinematics of a genuine three-point function is most conveniently parametrized in terms of two variables $z$ and $\zp$,
\beq
z\,\zp = u {\rm~~and~~} (1-z)\,(1-\zp)=v\,,
\eeq
or equivalently
\beq\label{eq:z_definition}
z= \frac{1}{2}\left(1+u-v+\sqrt{\lambda(1,u,v)}\right) {\rm~~and~~} \zp= \frac{1}{2}\left(1+u-v-\sqrt{\lambda(1,u,v)}\right)\,,
\eeq
where $\lambda$ denotes the K\"allen function
\beq
\lambda(a,b,c) = a^2+b^2+c^2-2ab-2ac-2bc\,.
\eeq
The appearance of the K\"allen function in eq.~\eqref{eq:z_definition} divides the $(u,v)$ plane into four different kinematical regions, shown in fig.~\ref{fig:regions}. In the regions II, III and IV the K\"allen function $\lambda(1,u,v)$ is positive, and so $z$ and $\zp$ are both real. In region I, on the contrary, we have $\lambda(1,u,v)<0$, and thus $z$ and $\zp$ are complex conjugate to each other. There is a fifth region where the K\"allen function vanishes, i.e., the boundary between region I and the regions II, II and IV. If we consider decay kinematics, region I is kinematically not allowed and has therefore often been discarded in the literature. In the following we take a different viewpoint and we argue that region I is the fundamental domain in which all three-mass triangle integrals are defined. The other regions are related to this fundamental domain by analytic continuation. In order to motivate this statement we need to introduce a mathematical tool that allows us to analyze the structure of polylogarithmic functions, the so-called symbol, which we briefly review in the rest of this section.

\subsection{Symbols of three-mass triangles and single-valued multiple polylogarithms}
Before defining the symbol map, we first need to define multiple polylogarithms, which are the natural set of functions describing large classes of Feynman integrals.
Multiple polylogarithms are a generalization of the ordinary logarithm and the \emph{classical} polylogarithms,
\beq\label{eq:classical_polylog}
\ln z = \int_1^z{\rd t\over t} {\rm~~and~~} \li_n(z) = \int_0^z{\rd t\over t}\,\li_{n-1}(t)\,,
\eeq
with $\li_1(z) = -\ln(1-z)$. It is known that multi-loop multi-scale integrals can give rise to new classes of functions, among which one finds the so-called \emph{multiple} polylogarithms, a multi-variable extension of eq.~\eqref{eq:classical_polylog} defined recursively via the iterated integral~\cite{Goncharov:1998, Goncharov:2001}
 \beq\label{eq:Mult_PolyLog_def}
 G(a_1,\ldots,a_n;z)=\,\int_0^z\,{\rd t\over t-a_1}\,G(a_2,\ldots,a_n;t)\,,\\
\eeq
with $G(z)=1$ and where $a_i, z\in \mathbb{C}$. In the special case where all the $a_i$'s are zero, we define, using the obvious vector notation $\vec a_n=(\underbrace{a,\dots,a}_{n})$,
\beq
G(\vec 0_n;z) = {1\over n!}\,\ln^n z\,.
\eeq
The number $n$ of elements $a_i$, counted with multiplicities, is called the {\em weight} of the multiple polylogarithm.
In special cases multiple polylogarithms can be expressed through classical polylogarithms and ordinary logarithms only, e.g.,
\beq\label{eq:G_reduce}
G(\vec a_n;z) = \frac{1}{n!}\ln^n\left(1-\frac{z}{a}\right) {\rm~~and~~} G(\vec0_{n-1},a;z) = -\li_{n}\left(\frac{z}{a}\right)\,.
\eeq
In general, however, no such simple formulae are known.
An important role is played in physics by the so-called \emph{harmonic} polylogarithms (HPLs)~\cite{Remiddi:1999ew}, which correspond, up to a sign, to the special case $a_i\in\{-1,0,1\}$,
\beq
H(a_1,\ldots,a_n;z) = (-1)^p\,G(a_1,\ldots,a_n;z)\,,
\eeq
where $p$ denotes the number of elements in $(a_1,\ldots,a_n)$ equal to $+1$. 

Iterated integrals form a {\em  shuffle algebra}~\cite{Ree:1958}, which allows one to express the product of two multiple polylogarithms of weight $n_1$ and $n_2$ as a linear combination with integer coefficients of multiple polylogarithms of weight $n_1+n_2$,
  \beq\bsp\label{eq:G_shuffle}
  G(a_1,\ldots,a_{n_1};z) \, G(a_{n_1+1},\ldots,a_{n_1+n_2};z) &\,=\sum_{\sigma\in\Sigma(n_1, n_2)}\,G(a_{\sigma(1)},\ldots,a_{\sigma(n_1+n_2)};z),\\
      \esp\eeq
where $\Sigma(n_1,n_2)$ denotes the set of all shuffles of $n_1+n_2$ elements, \emph{i.e.}, the subset of the symmetric group $S_{n_1+n_2}$ defined by
\beq\label{eq:Sigma_def}
\Sigma(n_1,n_2) = \{\sigma\in S_{n_1+n_2} |\, \sigma^{-1}(1)<\ldots<\sigma^{-1}({n_1}) {\rm~~and~~} \sigma^{-1}(n_1+1)<\ldots<\sigma^{-1}(n_1+{n_2})\}\,.
\eeq 

Multiple polylogarithms satisfy various complicated functional equations among themselves. A way to deal with the functional equations is given by the \emph{symbol}, a linear map which associates to every multiple polylogarithm an element in the tensor algebra over the group of rational functions. 
Various (equivalent) definitions have been given in the literature for the symbol of a multiple polylogarithm~\cite{Goncharov-simple-Grassmannian,symbolsC,symbolsB,Goncharov:2010jf,Duhr:2011zq,Duhr:2012fh}. One possible way to define the symbol of a multiple polylogarithm is to consider its total differential~\cite{Goncharov:2001},
\beq\label{eq:MPL_diff_eq}
dG(a_{n-1},\ldots,a_{1};a_n) = \sum_{i=1}^{n-1}G(a_{n-1},\ldots,\hat{a}_i,\ldots,a_{1};a_n)\, d\ln\left({a_i-a_{i+1}\over a_i-a_{i-1}}\right)\,,
\eeq
and to define the symbol recursively by~\cite{Goncharov:2010jf}
\beq\label{eq:GSVV_def}
\cS(G(a_{n-1},\ldots,a_{1};a_n)) = \sum_{i=1}^{n-1}\cS(G(a_{n-1},\ldots,\hat{a}_i,\ldots,a_{1};a_n))\otimes\left({a_i-a_{i+1}\over a_i-a_{i-1}}\right)\,.
\eeq
As an example, the symbols of the classical polylogarithms and the ordinary logarithms are given by
\beq
\cS(\li_{n}(z)) = -(1-z)\otimes\underbrace{z\otimes\ldots\otimes z}_{(n-1)\textrm{ times}} {\rm~~and~~} \cS\left(\frac{1}{n!}\ln^nz\right) = \underbrace{z\otimes\ldots\otimes z}_{n\textrm{ times}}\,.
\eeq
In addition the symbol satisfies the following identities,
\beq\bsp
\ldots\otimes (a\cdot b)\otimes\ldots &\,= \ldots\otimes a\otimes\ldots + \ldots\otimes b\otimes\ldots\,,\\
 \ldots\otimes(\pm1)\otimes\ldots &\,= 0\,.
\esp\eeq

The symbol also encodes information about the cuts of a function. As an example, let us take a function $f(z)$ whose symbol can be
written schematically in the form
\beq
\cS(f(z)) = a_1(z)\otimes \ldots\otimes a_n(z)\,,
\eeq
where $a_i(z)$ are rational functions of some variable $z$. Then $f(z)$ has branch cuts in the complex $z$ plane starting at points $z_i$ with $a_1(z_i)=0$ or $a_1(z_i) = \infty$, and the symbol of the
discontinuity across the branch cut is obtained by dropping the first entry in the symbol of $f(z)$\,,
\beq\label{eq:symb_disc}
\cS\left[\textrm{Disc}_{a_1(z_i)=0}f(z)\right] = a_2(z)\otimes\ldots\otimes a_n(z)\,.
\eeq
If $f(z)$ is a loop integral, then its cuts are determined by Cutkosky's rules. This implies in particular that the first entry of the symbol of a loop integral (with massless propagators)
must be a Mandelstam invariant~\cite{Gaiotto:2011dt}. In the special case of three-mass triangles with massless propagators we consider, this implies that at each order in the $\eps$ expansion,
\beq\label{eq:Laurent}
\mathcal{T}^{(\ell)}(u,v;\eps) = \sum_{k=m}^\infty\mathcal{T}^{(\ell)}_k(u,v)\,\eps^k\,,\qquad \textrm{for some } m\in\mathbb{Z}\,,
\eeq
the symbol of the coefficients of the Laurent series must take the form
\begin{eqnarray}\label{eq:triangle_symbol}
\cS\left[\mathcal{T}^{(\ell)}_k(u,v)\right] &=& \sum_i\left[R^u_i(u,v)\,u\otimes U_i(u,v) + R^v_i(u,v)\,v\otimes V_i(u,v)\right]\\
&=& \sum_i\left[\overline{R}^u_i(z,\zp)\,(z\zp)\otimes \overline{U}_i(z,\zp) + \overline{R}^v_i(z,\zp)\,[(1-z)(1-\zp)]\otimes \overline{V}_i(z,\zp)\right]\,,\nonumber
\end{eqnarray}
where $R_i^{u,v}$ are algebraic functions of the kinematic variables $u$ and $v$, and $U_i$ and $V_i$ are tensors of lower weight.

The form of the symbol~\eqref{eq:triangle_symbol} exhibits an important property of the functions $\mathcal{T}^{(\ell)}_k$. Let us assume that we are working in region  I
where $\lambda(1,u,v)<0$, and thus $z$ and $\zp$ are complex conjugate to each other. If we want to compute the discontinuities of the function $\mathcal{T}^{(\ell)}_k$ (seen as a function of $(z,\zp)$) in the complex $z$ plane, then it follows from eq.~\eqref{eq:symb_disc} that $\mathcal{T}^{(\ell)}_k$ has potential branch cuts starting at $z=0$, $z=1$ and $z=\infty$. We will concentrate for now on the discontinuity around $z=0$. The argument for the other two cases is similar. The symbol of $\textrm{Disc}_{z=0}\mathcal{T}^{(\ell)}_k$ only has contributions from the first term in eq.~\eqref{eq:triangle_symbol}. Terms of the form $z\otimes U_i$ and $\zp\otimes U_i$ obviously contribute with opposite signs, and so they cancel. The same argument holds for the discontinuities around $z=1$ and $z=\infty$. We thus conclude that the functions $\mathcal{T}^{(\ell)}_k$, seen as functions of the single complex variable $z$ and its complex conjugate $\zp$, have no branch cuts in the complex $z$ plane, and hence they are a single-valued functions of the complex variable $z$. We have thus shown that three-mass triangles without internal masses are expressible in region I through single-valued polylogarithms of the complex variable $z$. This is in agreement with the corresponding result for the generalized ladder integrals~\cite{Drummond:2012bg}, which reduce to three-mass triangle integrals upon using conformal invariance to send a point to infinity. 
We note however that while, strictly speaking, single-valuedness implies the first entry condition~\eqref{eq:triangle_symbol}, the inverse is not necessarily true. We discuss this problem in more detail in appendix~\ref{app:basis}. It is however possible to carry out a more careful analysis based on the Hopf algebra of multiple polylogarithms without altering the conclusion. The requirement that the functions be single-valued then strongly constrains the set of functions that can appear as coefficients in the Laurent series~\eqref{eq:Laurent}. This class of functions will be reviewed in the next section.

\section{Single-valued multiple polylogarithms in one variable}
\label{sec:SV_polylogs}
In this section we give a short review of single-valued polylogarithms in one complex variable $z$.
We start by defining the single-valued analogues of the classical polylogarithms. The classical
polylogarithms have branch cuts starting at $z=1$, and the discontinuity across the branch cut is given by
\begin{equation}\label{eq:lin_disc}
\textrm{Disc}_{z=1} {\rm Li}_n(z) = 2\pi i\, \frac{\log^{n-1} z}{(n-1)!}\,.
\end{equation}
The knowledge of the discontinuities~\eqref{eq:lin_disc} can be used to construct linear combinations
of classical polylogarithms in $z$ and its complex conjugate $\zp$ such that all the discontinuities cancel. 
As a result, we obtain a sequence of real-analytic functions on the punctured complex plane $\mathbb{C}/\{0,1\}$.
Although the space of single-valued functions is unique, there is a freedom of how to choose a basis in the space of single-valued functions, and several 
definitions of single-valued analogues of the classical polylogarithms have been proposed in the literature. Here we use the definition of Zagier, and we define
\beq\label{eq:P_n_def}
P_n(z) = \mathfrak{R}_n\left\{\sum_{k=0}^{n-1}\frac{2^k\,B_k}{k!}\,\ln^k|z|\,\li_{n-k}(z)\right\}\,,
\eeq
where $\mathfrak{R}_n$ denotes the real part if $n$ is odd and the imaginary otherwise, and $B_k$ are the Bernoulli numbers, defined as the coefficients in the Taylor expansion
\beq
\frac{z}{e^{z}-1} = \sum_{k=0}^\infty B_k\,\frac{z^k}{k!}\,.
\eeq
The functions $P_n(z)$ are known to satisfy nice functional equations, e.g.,
\beq\label{eq:Pn_inversion}
P_n\left(\frac{1}{z}\right) = (-1)^{n+1}\,P_n(z)\,,\qquad n\ge2\,.
\eeq
For $n=2$, eq.~\eqref{eq:P_n_def} reduces to the famous Bloch-Wigner dilogarithm, which satisfies in addition the functional equation
\beq\label{eq:P2_reflection}
P_2(1-z) = -P_2(z)\,.
\eeq

Just like it is possible to define single-valued analogues of the classical polylogarithms, there have been various attempts to define single-valued versions of multiple polylogarithms. In particular, single-valued versions have been constructed for all multiple polylogarithms up to weight three~\cite{Zhao} as well as for all harmonic polylogarithms with indices $a_i\in\{0,1\}$~\cite{BrownSVHPLs}. The single-valued harmonic polylogarithms are characterized by symbols of the form~\eqref{eq:triangle_symbol}, where $U_i$ and $V_i$ are lower weight tensors whose entries are drawn from the set $\{z,\zp,1-z,1-\zp\}$. Furthermore, they can always be written in the factorized form
\beq\label{eq:factorized_form}
\mathcal{L}_w(z) = \sum_{i,j} c_{ij}\,H(\vec a_i;z)\, H(\vec a_j;\zp)\,,
\eeq
where the coefficients $c_{ij}$ are rational polynomials in multiple zeta values, and $\vec a_i$ and $\vec a_j$ are (possibly empty) sequences of 0's and 1's.
It is clear from eq.~\eqref{eq:triangle_symbol} that single-valued HPLs provide a natural subset of the possible functions that can appear in the $\eps$ expansion of three-mass triangles. However, it is known that up to weight four all harmonic polylogarithms with indices 0 and 1 are expressible in terms of classical polylogarithms only~\cite{Duhr:2011zq}. As in this paper we are only interested in two-loop integrals for which polylogarithms of weight at most four can appear, we can restrict ourselves to the single-valued versions of the classical polylogarithms given in eq.~\eqref{eq:P_n_def}.

As we will see in subsequent sections, the single-valued versions of the classical polylogarithms are however not sufficient to write down all three-mass triangle integrals at two loops. Indeed, our results show that the symbols of three mass triangles do not only have entries drawn from the set $\{z,\zp,1-z,1-\zp\}$, but also the entry $z-\zp$ appears. We therefore conjecture that, to all loop orders and to all orders in $\eps$, three mass triangles are linear combinations of functions whose symbols
\begin{enumerate}
\item have entries drawn from the set $\{z,\zp,1-z,1-\zp,z-\zp\}$,
\item have a first entry which is either $(z\zp)$ or $(1-z)(1-\zp)$.
\end{enumerate}
Our conjecture implies that, for each weight, we can predict a priori the space of possible transcendental functions that can appear in the answer. In particular, we can construct a basis of functions at each weight, and in practice this basis turns out to be rather small. The indecomposable basis elements of a given weight (i.e., basis elements which cannot be written as linear combinations of products of lower weights) contain a set of indecomposable single-valued harmonic polylogarithms~\cite{BrownSVHPLs}, augmented by some new functions whose symbols contain an entry equal to $z-\zp$. To our knowledge, these new functions have never been studied in the literature so far. In appendix~\ref{app:basis} we therefore present a recursive construction for the new basis elements. As an example, the only new single-valued function of weight three besides single-valued HPLs is
\beq\bsp\label{eq:Q3_def}
\mathcal{Q}_3(z)&\, = \frac{1}{2} \left[G\left(0,\frac{1}{\zp},\frac{1}{z},1\right)-G\left(0,\frac{1}{z},\frac{1}{\zp},1\right)\right]+\frac{1}{4} \log |z|^2 \left[G\left(\frac{1}{z},\frac{1}{\zp},1\right)-G\left(\frac{1}{\zp},\frac{1}{z},1\right)\right]\\
&\,+\frac{1}{2} \Big[\text{Li}_3(1-z)-\text{Li}_3(1-\zp)\Big]+\text{Li}_3(z)-\text{Li}_3(\zp)+\frac{1}{4} \Big[\text{Li}_2(z)+\text{Li}_2(\zp)\Big] \log \frac{1-z}{1-\zp}\\
&\,+\frac{1}{4} \Big[\text{Li}_2(z)-\text{Li}_2(\zp)\Big] \log |1-z|^2+\frac{1}{16} \log \frac{z}{\zp} \log^2 \frac{1-z}{1-\zp}+\frac{1}{8} \log^2 |z|^2 \log \frac{1-z}{1-\zp}\\
&+\frac{1}{4} \log|z|^2\,\log|1-z|^2\, \log \frac{1-z}{1-\zp}+\frac{1}{16} \log^2|1-z|^2\log\frac{z}{\zp}-\frac{\pi ^2}{12}  \log \frac{1-z}{1-\zp}\,.
\esp\eeq
Similar results for the new basis functions of weight four can be found in appendix~\ref{app:basis}. 
The main difference between the new basis functions and the single-valued HPLs is that the new functions cannot be written in a  factorized form~\eqref{eq:factorized_form}, but they involve genuine multiple polylogarithms in $(z,\zp)$. The proof that $\cQ_3(z)$ is indeed single-valued in the complex $z$ plane follows from the construction of appendix~\ref{app:basis}. We note that, as every multiple polylogarithm of weight at most three can be expressed through classical polylogarithms only, we could derive an expression for $\cQ_3(z)$ that does not involve any multiple polylogarithm. The result would be a combination of classical polylogarithms which individually have a very complicated branch cut structure, and the different cuts conspire such that $\cQ_3(z)$ is single-valued. We therefore prefer to present $\cQ_3(z)$ in the form~\eqref{eq:Q3_def}. Furthermore note that, just like the singe-valued analogues of the classical polylogarithms~\eqref{eq:P_n_def}, $\mathcal{Q}_3(z)$ has a definite parity under complex conjugation, $z\leftrightarrow\zp$. More generally, we can choose all the basis elements as eigenstates of the action of the $\mathbb{Z}_2$ symmetry group corresponding to complex conjugation.
The indecomposable basis elements up to weight four with given parity under complex conjugation are shown in tab.~\ref{tab:indecomposables}. Note that we introduce the short-hand
\beq
\cP_n(z) \equiv \left\{\begin{array}{ll}
2 P_n(z)\,,&\textrm{ if } n \textrm{ odd}\,,\\
2 i P_n(z)\,,&\textrm{ if } n \textrm{ even}\,,\\
\end{array}\right.
\eeq
in order to absorb the normalization factor coming from the real and imaginary part (because we will have to consider these functions as well in the region II, III, IV, where $z$ and $\zp$ are not complex conjugate to each other).
\begin{table}[!t]
\begin{center}
\begin{tabular}{c|c|c}
\hline\hline
weight & + & -\\
\hline
1 & $\displaystyle\ln|z|^2, \ln|1-z|^2$ & -- \\
2 & $\zeta_2$ & $\displaystyle\cP_2(z)$ \\
3 & $\zeta_3, \cP_3(z), \cP_3(1-z)$ & $\displaystyle\cQ_3(z)$ \\
4 &  $\displaystyle\cQ^+_4(z), \cQ^+_4(1-z)$& $\displaystyle\cP_4(z), \cP_4(1-z), \cP_4(1-1/z), \cQ_4^-(z)$ \\
\hline\hline
\end{tabular}
\caption{\label{tab:indecomposables}Indecomposables basis elements up to weight four which can appear in the $\epsilon$ of three mass triangle integrals.}
\end{center}
\end{table}

Let us conclude this section with a discussion on how the symmetries of three-mass triangle integrals are implemented into the space of single-valued polylogarithms we just defined.
There is a natural action of the symmetric group $S_3$ on three-mass triangles, acting by a permutation of the external legs. This $S_3$ symmetry acts on the space of single-valued functions via 
\begin{eqnarray}\label{eq:S3_symmetry}
z\to z \,, \qquad & z \to 1-\zp\,, \qquad &z \to 1-1/z \,,\\
z\to 1/\zp\,,\qquad &z\to 1/(1-z)\,,\qquad &z\to \zp/(\zp-1)\,.\nonumber
\end{eqnarray}
It is easy to check that the space of single-valued functions we just defined is closed under these transformations,
because the set of irreducible polynomials $\{z,\zp,1-z,1-\zp,z-\zp\}$ is invariant under the transformations~\eqref{eq:S3_symmetry} and also the first entry condition is preserved. 
As a consequence, the symmetries of a three-mass triangle graph are implemented into the space of functions via the functional equations arising from the transformations
of the argument~\eqref{eq:S3_symmetry}. We have worked out the relevant functional equations for all the basis functions up to weight four. The results are shown in appendix~\ref{app:basis}.

\section{Three-mass triangles in region I}
\label{sec:generic_z}
\subsection{The one-loop triangle}
\label{sec:oneloop}
In this section we compute the analytic expressions of the two-loop three-mass triangle master integrals in region I where $\lambda(1,u,v)<0$. The results for the other regions will be given in subsequent sections.

As a warm-up, we consider the one-loop three-mass triangle and compute its Laurent expansion in $\eps$ up to terms in the expansion of transcendental weight four. More precisely, we consider the integral
\beq
T_1(p_1^2,p_2^2,p_3^2;\eps) = e^{\gamma_E\eps}\,\int{\rd^Dk\over i\pi^{D/2}}\,{1\over k^2\,(k+p_1)^2\,(k-p_2)^2}\,,
\eeq
with $p_i^2\neq 0$ and $D=4-2\eps$. After Feynman parametrization we obtain
\beq
T_1(p_1^2,p_2^2,p_3^2;\eps) = - c_\Gamma\,{\Gamma(1-2\eps)\over\Gamma(1-\eps)^2}\,(-p_3^2)^{-1-\eps}\,\cI(1,1,1;u,v;\eps)\,,
\eeq
where we defined
\beq\bsp\label{eq:I_definition}
\cI(\nu_1,\nu_2,\nu_3;u,v;\eps) =&\, \int_0^\infty \left(\prod_{k=1}^3\rd x_i\,\frac{x_1^{\nu_i-1}}{\Gamma(\nu_i)}\right)\,\delta\left(1-\sum_{i\in S}x_i\right)\\
&\qquad\times(x_1+x_2+x_3)^{\nu-D}\,(x_2\,x_3+x_1\,x_2\,u+x_1\,x_3\,v)^{D/2-\nu}\,,
\esp\eeq
with $\nu=\nu_1+\nu_2+\nu_3$ and $S$ any non-empty subset of $\{1,2,3\}$~\cite{Cheng:1987ga}. In the following we choose $S=\{1\}$ and we obtain
\beq\bsp\label{eq:I_int}
\cI(\nu_1,\nu_2,&\nu_3;u,v;\eps) \\
&= \int_0^\infty \rd x_2\,\rd x_3 \,x_2^{\nu_2-1}\,x_3^{\nu_3-1}\,(1+x_2+x_3)^{\nu-D}\,(x_2\,x_3+x_2\,u+x_3\,v)^{D/2-\nu}\,.
\esp\eeq
The integral $\cI(1,1,1;u,v;\eps)$ is finite as $\eps\to0$ and can thus be expanded in $\eps$ under the integration sign. We perform the integration order by order in $\eps$ using the method of ref.~\cite{Brown:2008um,symbolsB,DelDuca:2009ac}, i.e., we perform the Feynman parameter integrals one by one using the following recursive procedure:
\begin{enumerate}
\item Choose a Feynman parameter $x_i$ and write all polylogarithms in the integrand as multiple polylogarithms of the form $G(\vec a; x_i)$, where $\vec a$ is a function of the remaining Feynman parameters and/or the external parameters. 
\item Use the shuffle algebra of multiple polylogarithms to replace every product of multiple polylogarithms in $x_i$ by a linear combination of such functions.
\item If all the denominators in the integral are linear, partial fraction and compute a primitive using the definition of multiple polylogarithms, eq.~\eqref{eq:Mult_PolyLog_def}.
\item Compute the value of the primitive at the boundaries of the integration region, and return to 1.
\end{enumerate}
Some comments are in order about this procedure: First, in order for the algorithm to converge, it is necessary to find at each step a Feynman parameter in which all the denominators are linear\footnote{Such integrals are called \emph{Fubini reducible} in ref.~\cite{Brown:2008um}.}. This condition is always satisfied in our case, as the integral~\eqref{eq:I_definition} is two-dimensional and involves only denominators that are linear in each Feynman parameter. Second, the first step of rewriting all the polylogarithms in the integrand in the form $G(\vec a; x_i)$ involves the use functional equations among multiple polylogarithms. This task can easily be carried out by using the Hopf algebra structure of multiple polylogarithms, which allows to derive complicated functional equations among multiple polylogarithms in an easy way. Finally we have to address the question of how to take the limit $x_i\to\infty$ of the primitive. We will illustrate this step on the example below.

Using the algorithm we just described, we can easily obtain, at least in principle, the $\eps$ expansion of the one-loop triangle to any order. We will illustrate this procedure in detail on the example of the coefficient of $\eps^0$. In this case we are left with the integral
\beq\bsp
\cI(1,1,&1;u,v;0) = \int_0^\infty\frac{\rd x_2\,\rd x_3}{(1+x_2+x_3)\,(x_2\,x_3+x_2\,u+x_3\,v)}\,.
\esp\eeq
The integral over $x_2$ is trivial to perform and we get
\beq\label{eq:I_x3}
\cI(1,1,1;u,v;0) =\int_0^\infty \rd x_3\,
\frac{\log \left(1+x_3\right)-\ln v+\log \left(u+x_3\right)-\log x_3}{x_3^2+(1+u-v)\, x_3+v}\,.
\eeq
The roots of the quadratic denominator are given by $(-z,-\zp)$ defined in eq.~\eqref{eq:z_definition}. 
While in the present case the integration over $x_3$ could easily be carried out in terms of dilogarithms, we prefer nevertheless to compute it explicitly using the algorithm outlined above in order to illustrate its usage for the higher-order terms in the $\eps$ expansion. Using eq.~\eqref{eq:G_reduce} the integral can be recast in the form
\beq
\cI(1,1,1;u,v;0) =\int_0^\infty \rd x_3\,
\frac{G(-1;x_3)+\ln u-\ln v + G(-u;x_3)-G(0;x_3)}{(x_3+z)(x_3+\zp)}\,.
\eeq
Note that if $\lambda(1,u,v)<0$, the integral is manifestly real and convergent for all positive values of $u$ and $v$. Using eq.~\eqref{eq:Mult_PolyLog_def}, it is easy to find a primitive of the integrand,
\beq\bsp
\rho(z,\zp;x_3) = \frac{1}{z-\zp}\Big\{&
-[G(-z,-u;x_3)-G(-\zp,-u;x_3)]-\log u [G(-z;x_3)-G(-\zp;x_3)]\\
&+\log v [G(-z;x_3)-G(-\zp;x_3)]-[G(-z,-1;x_3)-G(-z,0,x_3)]\\
&-[G(-\zp,0;x_3)-G(-\zp,-1;x_3)]\Big\}\,.
\esp\eeq
Next we have to compute the limits $x_3\to0$ and $x_3\to\infty$ of the primitive. The first limit is trivial, and we obtain
\beq\label{eq:limit}
\cI(1,1,1;u,v;0) = \lim_{x_3\to\infty}\rho(z,\zp;x_3)\,.
\eeq
In order to take the limit at infinity, we have to use the inversion relations, $x_3\to1/x_3$, for the multiple polylogarithms appearing inside the primitive. These identities can easily be derived using the Hopf algebra of multiple polylogarithms. Indeed, using the techniques developed in refs.~\cite{Duhr:2011zq, Duhr:2012fh} it is easy to show that the following identity holds for $x_3>0$ and $z$ a generic complex number,
\beq\bsp
\rho(z,\zp;x_3) = \frac{1}{z-\zp}\Bigg\{&-G\left(-\frac{1}{z},-\frac{1}{u};\frac{1}{x_3}\right)+G\left(-\frac{1}{\zp},-\frac{1}{u};\frac{1}{x_3}\right)-\log u [G(u;z)-G(u;\zp)]\\
&+G(u,0;z)-G(u,0;\zp)+\log v \left[G\left(-\frac{1}{z};\frac{1}{x_3}\right)-G\left(-\frac{1}{\zp};\frac{1}{x_3}\right)\right]\\
&-G\left(-\frac{1}{z},-1;\frac{1}{x_3}\right)+G\left(-\frac{1}{z},0;\frac{1}{x_3}\right)+\text{Li}_2(z)-\text{Li}_2(\zp)\\
&+G\left(-\frac{1}{\zp},-1;\frac{1}{x_3}\right)-G\left(-\frac{1}{\zp},0;\frac{1}{x_3}\right)+\log u (\log z-\log \zp)\\
&-\log v (\log z-\log \zp)-\frac{1}{2} [\log ^2z-\log^2\zp]+\log (1-z) \log z-\log (1-\zp) \log \zp\Big\}\,.
\esp\eeq
The limit~\eqref{eq:limit} is now trivial to take, and we obtain
\beq\bsp\label{eq:result_oneloop_raw}
\cI(1,1,1;u,v;0) = \frac{1}{z-\zp}\Bigg\{&
-\log u [G(u;z)-G(u;\zp)]+G(u,0;z)-G(u,0;\zp)\\
&+\text{Li}_2(z)-\text{Li}_2(\zp)+\log u [\log z-\log \zp]-\log v [\log z-\log \zp]\\
&-\frac{1}{2} [\log ^2z-\log ^2\zp]+\log (1-z) \log z-\log (1-\zp) \log \zp\Bigg\}\,.
\esp\eeq
Eq.~\eqref{eq:result_oneloop_raw} is the correct expression for the three-mass one-loop triangle in four dimensions. 
However, in section~\ref{sec:SV_polylogs} we argued that all three-mass triangle up to two loops can be expressed in region I through single-valued functions only. 
The polylogarithms appearing in $\cI(1,1,1;u,v;0)$ are manifestly odd under $z\leftrightarrow\zp$, and tab.~\ref{tab:indecomposables} shows that there is precisely one odd basis element of weight two. Indeed, if we compute the symbol of eq.~\eqref{eq:result_oneloop_raw} we get
\beq
\cS\Big[\cI(1,1,1;u,v;0)\Big] = \frac{1}{z-\zp}\Big\{(z \zp)\otimes \frac{1-z}{1-\zp}-[(1-z) (1-\zp)]\otimes \frac{z}{\zp}\Big\} = \frac{1}{z-\zp}\,\cS[2\,\cP_2(z)]\,.
\eeq
Note that we cannot add any rational multiple of $\zeta_2$ to the argument of the symbol in the right-hand side because of the parity of the function. We thus arrive at the conclusion that
\beq
\cI(1,1,1;u,v;0) = \frac{2}{z-\zp}\,\cP_2(z)\,.
\eeq

Our strategy can immediately be extended to the higher-order terms in the Laurent expansion. 
The only technical difficulty that arises is that the $x_2$ integration is no longer trivial, but the integrand now involves powers of logarithms of the denominators. For example, the coefficient of $\eps^n$ is given by the integral
\beq\bsp
\sum_{k=0}^n (-1)^{n-k}\,2^k\,\binom{n}{k}\int_0^\infty&\frac{\rd x_2\,\rd x_3}{(1+x_2+x_3)\,(x_2\,x_3+x_2\,u+x_3\,v)}\\
&\times\log^k(1+x_2+x_3)\,\log^{n-k}(x_2\,x_3+u\,x_2+v\,x_3)\,.
\esp\eeq
The logarithms can be written as multiple polylogarithms in $x_2$,
\beq\bsp
\log^k&(1+x_2+x_3)\,\log^{n-k}(x_2\,x_3+u\,x_2+v\,x_3) \\
&= \sum_{p=0}^k\sum_{q=0}^{n-k}\binom{k}{p}\,\binom{n-k}{q}\,\log^{k-p}(1+x_3)\,\log^{n-k-q}(v\,x_3)\,\\
&\qquad\qquad \times G\left(-1-x_3;x_2\right)^p\,G\left(-\frac{v\,x_3}{u+x_3};x_2\right)^q\,.
\esp\eeq
The products of multiple polylogarithms can be linearized using the shuffle algebra, and we can thus rewrite the integrand in terms of rational functions and multiple polylogarithms in $x_2$. If we treat $x_3$ as a constant, this integral can be performed in exactly the same way as the integral over $x_3$ discussed previously.

Carrying out this procedure for the first few terms in the $\eps$ expansion,
 we find
\beq\bsp\label{eq:one_loop_triangle}
T_1(p_1^2,p_2^2,p_3^2;\eps) = &\,-2 c_\Gamma\,{\Gamma(1-2\eps)\over\Gamma(1-\eps)^2}\,(-p_3^2)^{-1-\eps}\,{u^{-\eps}\,v^{-\eps}\over z-\zp}\,
\Bigg\{\cP_2(z) + 2\eps\cQ_3(z)\\
& \,+ \eps^2\,\Big[
\Big(\frac{1}{6} \log u \log v -\zeta_2\Big)\cP_2(z)+2\, \cQ_4^-(z)\Big]+\ord(\eps^3)\Bigg\}\,.
\esp\eeq
As anticipated in section~\ref{sec:SV_polylogs}, the result for the one-loop three-mass triangle is expressed, at each order in $\eps$, as a combination of the single-valued polylogarithms shown in table~\ref{tab:indecomposables}. We stress that the functions appearing in eq.~\eqref{eq:one_loop_triangle} form a basis of the space of functions in which the mass triangles naturally take values. Thus, this expression is minimal, and there are no further relations among these functions. 

While this is not the first time that an analytic expression for the one-loop three-mass triangle has been computed, we believe that our result is an improvement over existing representations given in the literature. In refs.~\cite{Usyukina:1994iw,tHooft} analytic expressions for the one-loop three-mass triangle in $D=4$ dimensions in terms of dilogarithms were computed. However, in the computation of NNLO observables also the higher order terms in the $\eps$ expansion were presented. In ref.~\cite{Davydychev:1999mq} the one-loop three-mass triangle was expressed in region I, to all orders in $\eps$ in terms of log-sine integrals, while in ref.~\cite{Birthwright:2004kk} the same function was expressed up to $\ord(\eps^2)$ in region II in terms of complicated iterated integrals whose kernels involve inverse square roots of the K\"allen function. These representations of the function are only valid in specific regions in $(u,v)$ space, and the analytic continuation from one region to another can be very complicated. While it appears that our derivation makes explicit use of the fact that we work in region I, we will see in the next section that the analytic continuation of eq.~\eqref{eq:one_loop_triangle} to all other regions is straightforward. 

Let us conclude this section by a comment about the symmetries of the one-loop three-mass triangle. Indeed, in section~\ref{sec:SV_polylogs} we argued that the $S_3$ symmetry permuting the external legs of the triangle graph are encoded into the space of functions via the functional equations corresponding to the transformations of the argument~\eqref{eq:S3_symmetry}. It is easy to check using the functional equations for the basis functions, eqs.~\eqref{eq:P2_reflection} and~\eqref{eq:Pn_inversion}, as well as the functional equations of appendix~\ref{app:functional_equations}, that our result is invariant under $z\to1-\zp$ and $z\to 1/\zp$. As these two transformations generate the whole symmetric group $S_3$, eq.~\eqref{eq:one_loop_triangle} is invariant under the whole action of $S_3$, as expected.

\subsection{The two-loop master integrals}
\label{sec:two_loop_MI}
After this warm-up, we turn in this section to the two-loop master integrals shown in fig.~\ref{fig:master_integrals}.
The master integrals relevant for two-loop three-point off-shell functions were identified in ref.~\cite{Birthwright:2004kk},
\begin{eqnarray}
F_1(p_1^2,p_2^2,p_3^3;\eps) &= &e^{2\gamma_E\eps}\,\int{\rd^Dk\,\rd^D\ell\over (i\pi^{D/2})^2}\,{1\over (k+p_1)^2\,(k-p_2)^2\,\ell^2\,(k-\ell)^2}\,,\nonumber\\
F_2(p_1^2,p_2^2,p_3^3;\eps) &= &e^{2\gamma_E\eps}\,\int{\rd^Dk\,\rd^D\ell\over (i\pi^{D/2})^2}\,{1\over (k+p_1)^2\,(k-p_2)^2\,\ell^2\,[(k-\ell)^2]^2}\,,\\
F_3(p_1^2,p_2^2,p_3^3;\eps) &= &e^{2\gamma_E\eps}\,\int{\rd^Dk\,\rd^D\ell\over (i\pi^{D/2})^2}\,{1\over k^2\,(k+p_1)^2\,\ell^2\,(k+p_1-\ell)^2\,(k-p_3-\ell)^2\,}\,,\nonumber\\
F_4(p_1^2,p_2^2,p_3^3;\eps) &= &e^{2\gamma_E\eps}\,\int \frac{\rd^Dk\,\rd^D\ell}{(i\pi^{D/2})^2}\, \frac{1}{k^2\ell^2(k+p_1)^2(l+p_2)^2(k-l+p_1)^2(l-k+p_2)^2}\,.\nonumber
\end{eqnarray}
The master integral $F_3$ was already computed in ref.~\cite{Usyukina:1992jd},
\beq\label{eq:F3_result}
F_3(p_1^2,p_2^2,p_3^3;\eps)=c_\Gamma^2\,(-p_3^2)^{-1-2\eps}\,{6\over z-\zp}\,\cP_4\left(1-{1\over z}\right) + \ord(\eps)\,.
\eeq
Similarly, it was shown in ref.~\cite{Usyukina:1994iw} that the non-planar triangle in four dimension is given by the square of the one-loop triangle,
\beq\label{eq:F4_result}
F_4(p_1^2,p_2^2,p_3^3;\eps) = 4\,c_\Gamma^2\,{(-p_3^2)^{-2-2\eps}\over (z-\zp)^2}\,\cP_2(z)^2+\ord(\eps)\,.
\eeq
As expected the integrals can be expressed in terms of single-valued polylogarithms. Furthermore, the symmetries of the diagrams are again entirely encoded into the functional 
equations among the basis functions. Indeed, it is easy to see from eqs.~\eqref{eq:Pn_inversion} and~\eqref{eq:P2_reflection} that $F_4$ is invariant under $S_3$ transformations. The planar integral $F_3$ is only invariant under a $\mathbb{Z}_2$ subgroup acting via permutation of the external legs $p_1$ and $p_2$, or equivalently $z\leftrightarrow1-\zp$. From the inversion relation~\eqref{eq:Pn_inversion} and the fact that $\cP_4(z)$ is odd under $z\leftrightarrow\zp$ we immediately see that eq.~\eqref{eq:F3_result} has the required symmetry,
\beq
\cP_4\left(1-{1\over z}\right) \to\cP_4\left(1-{1\over 1-\zp}\right)=-\cP_4\left({z\over z-1}\right)=\cP_4\left(1-{1\over z}\right)\,.
\eeq

Next we turn to the computation of the two remaining master integrals $F_1$ and $F_2$, which correspond to one-loop triangles with a bubble insertion.
While these integrals have already been considered in the literature, we believe that our results are genuinely new and constitute an improvement over the existing representations. Indeed, in ref.~\cite{Usyukina:1994iw} analytic expressions for $F_1$ and $F_2$ were given in terms of classical polylogarithms up to weight three. It is however known that two-loop computations in general require polylogarithms up to weight four, and thus an additional order in $\eps$ is required. This task was already performed in ref.~\cite{Birthwright:2004kk} where expressions for $F_1$ and $F_2$ were derived in terms of complicated iterated integrals  of weight four which are not obviously related to multiple polylogarithms. In the following we present analytic expressions for $F_1$ and $F_2$ which are given entirely in terms of multiple polylogarithms up to weight four.

We start by integrating out the bubble subintegral in each of the two master integrals using the formula,
\beq
\int{\rd^D\ell\over i\pi^{D/2}}\,{1\over[-\ell^2]^{\nu_1}\,[-(k-\ell)^2]^{\nu_2}} = (-k^2)^{\frac{D}{2}-\nu_1-\nu_2}\,\frac{\Gamma(\nu_1+\nu_2-\frac{D}{2})\,\Gamma(\frac{D}{2}-\nu_1)\,\Gamma(\frac{D}{2}-\nu_2)}{\Gamma(\nu_1)\,\Gamma(\nu_2)\,\Gamma(D-\nu_1-\nu_2)}\,.
\eeq
As a result, we can write each of the two master integrals as a one-loop triangle with one of the propagators raised to an $\eps$-dependent exponent. We obtain
\beq\bsp\label{eq:F_kappa}
F_{1+\kappa}&(p_1^2,p_2^2,p_3^2;\eps) \\
&\,={c_\Gamma^2\over (2-\kappa)\,\eps(1-2\delta_{0,\kappa}\eps)}\,{\Gamma(1+2\eps)\,\Gamma(1-2\eps)\over \Gamma(1+\eps)^2\,\Gamma(1-\eps)^2}\,(-p_3^2)^{-\kappa-2\eps}\,\cI(\kappa+\eps,1,1;u,v;\eps)\,,
\esp\eeq
where $\kappa\in\{0,1\}$ and $\cI(\nu_1,\nu_2,\nu_3;u,v;\eps)$ is the Feynman-parametrized one-loop triangle given in eq.~\eqref{eq:I_int}. As a consequence we can apply our strategy of section~\ref{sec:oneloop} to compute the master integrals $F_1$ and $F_2$. We will however deal with the two integrals separately as they have a different singularity structure.

Let us start by considering the master integral $F_2$, which corresponds to $\kappa=1$. Even though eq.~\eqref{eq:F_kappa} has an explicit pole in $\eps$, the integral $\cI(1+\eps,1,1;u,v;\eps)$ is finite as $\eps\to0$. We can thus expand under the integration sign and apply the strategy of section~\ref{sec:oneloop}. We obtain
\beq\bsp
\cI(1+\eps,1,1;u,v;\eps)
=\left(\frac{u^{-\eps}+v^{-\eps}}{2} -\eps^2\,\zeta_2\right)\cI(1,1,1;u,v;\eps) + \frac{6\,\epsilon^2}{z-\zp}\,\cP_4\left(1-{1\over z}\right) +\ord(\eps^3)\,.
\esp\eeq
We observe that the integral $F_2$ can be expressed as a linear combination of the one-loop triangle and the master integral $F_3$,
\beq\bsp\label{eq:F2}
F_{2}(p_1^2,p_2^2,p_3^2;\eps) &\,= -\frac{c_\Gamma}{\eps}\,{\Gamma(1+2\eps)\over \Gamma(1+\eps)^2}\,(-p_3^2)^{-\eps}\,\left(\frac{u^{-\eps}+v^{-\eps}}{2} -\eps^2\,\zeta_2\right)\,T_1(p_1^2,p_2^2,p_3^2;\eps)\\
&\,+\eps\,{\Gamma(1+2\eps)\,\Gamma(1-2\eps)\over \Gamma(1+\eps)^2\,\Gamma(1-\eps)^2}\,F_{3}(p_1^2,p_2^2,p_3^2;\eps)+\ord(\eps^2)\,.
\esp\eeq
So far we have no explanation for this decomposition of $F_2$ into two other master integrals.

Next we turn to the computation of the master integral $F_1$. Unlike the previous case, the integral $\cI(\eps,1,1;u,v;\eps)$ diverges as $\eps\to0$, and so we cannot expand under the integration sign. Instead we perform the integral over $x_3$ for finite $\eps$ in terms of hypergeometric functions,
\beq\bsp
\cI&(\eps,1,1;u,v;\eps) = \int_0^\infty\rd x_2\Bigg\{\frac{\Gamma (1-2 \epsilon ) \Gamma (1-\epsilon )}{\Gamma (2-3 \epsilon )} \frac{\left(v+x_2\right)^{1-3 \epsilon }}{\left[\left(1+x_2\right) \left(v+x_2\right)-u x_2\right]^{1-\eps}}\\
&-\frac{u^{1-2 \epsilon } x_2^{1-2 \epsilon } \left(1+x_2\right)^{3 \epsilon -2} \, }{(1-2 \epsilon ) \left(v+x_2\right)}\,{}_2F_1\left(1,2-3 \epsilon, 2-2 \epsilon ;\frac{u x_2}{\left(1+x_2\right) \left(v+x_2\right)}\right)\Bigg\}\,.
\esp\eeq
Note that in region  I the argument of the hypergeometric function is always less than 1. The divergence in the integral comes entirely from the term in the first line. It is easy to write down a subtraction term which renders the integral finite,
\beq
Z(x_2;\eps) = \frac{ \Gamma (1-2 \epsilon )\, \Gamma (1-\epsilon )}{\Gamma (2-3 \epsilon )}\,\left(1+x_2\right)^{-1-\epsilon}\,,
\eeq
and
\beq
\int_0^\infty\rd x_2\,Z(x_2;\eps) = \frac{\Gamma (1-2 \epsilon )\, \Gamma (1-\epsilon )}{\epsilon\,  \Gamma (2-3 \epsilon )}\,.
\eeq
After subtraction of the divergence, we can again expand under the integration sign and integrate out the Feynman parameters one-by-one. The only technical difficulty is that individual terms in the result have logarithmic divergences which cancel in the final answer. At the end of this procedure we find
\beq\bsp\label{eq:I1}
\cI&(\eps,1,1;u,v;\eps) = \frac{1}{1-3\eps}\,\Bigg\{\frac{1}{\eps} + \eps(1-u-v)\,\cI(1+\eps,1,1;u,v) - \eps\left[2\zeta_2 + \ln v\ln u\right] \\ 
&+ \eps^2\left[-6\,\cP_3(z)-6\,\cP_3(1-z) + \frac{1}{2}\ln u\ln^2 v + \frac{1}{2}\ln^2 u \ln v + 6\,\zeta_3 \right] \\ 
&+ \eps^3\Big[\frac{3}{4}\,\cQ_4^+(z) + \frac{3}{4}\,\cQ_4^+(1-z) - \frac{3}{2}(3\ln u - \ln v)\cP_3(z) + \frac{3}{2}(\ln u - 3\ln v)\cP_3(1-z)\\
& -\frac{9}{2}\,\cP_2(z)^2 + \frac{9}{16}(\ln^4 u + \ln^4 v) - \frac{25}{24}\,\ln u\,\ln v\,(\ln^2 u + \ln^2 v) +\frac{3}{8}\ln^2 u\ln^2 v\\
& -7\,\zeta_2\,\ln u\ln v -12\,\zeta_3\,\ln(uv)+7\zeta_4 \Big] \Bigg\} +\ord(\eps^4)\,.
\esp\eeq
We see that the integral is composed of two pieces: one piece contains the master integral $F_2$ multiplied by an algebraic prefactor, while the second part is made of single-valued polylogarithms that are even under $z\leftrightarrow\zp$. This is in agreement with the result obtained in ref.~\cite{Usyukina:1994iw} up to $\ord(\eps)$. Furthermore, we note that the function $\cQ_4^+(z)$ cannot be expressed through classical polylogarithms only, thus proving that $F_1$ must necessarily contain multiple polylogarithms. We refer to appendix~\ref{app:basis} for a more detailed discussion of this matter.

So far all our results are valid only in region I where $\lambda(1,u,v)<0$. In this region all the functions are expressed through single-valued functions in the complex variable $z$. 
In the next section we show how our results can be analytically continued to all the other regions shown in fig.~\ref{fig:regions}.


\section{Analytic continuation of three-mass triangle integrals}
\label{sec:anal_cont}
In the previous section we presented analytic results for all one and two-loop master integrals for three-mass triangle integrals valid in region I where $\lambda(1,u,v) < 0$.
In applications it is however important to know the expressions of the master integrals also in the other regions. In this section we perform the analytic continuation into the regions II, III and IV where $\lambda(1,u,v) >0$, and also into the boundary region V where $\lambda(1,u,v) =0$.

\subsection{Analytic continuation to the regions II, II and IV}
Let us start by by analytically continuing all the master integrals from region I where the K\"allen function is negative to the regions II, III and IV where the K\"allen function is positive (but non zero). The main difference with respect to region I is that $z$ and $\zp$ are real, and hence they are not complex conjugate to each other anymore. This implies that we will need a prescription of how to deal with polylogarithms that develop an imaginary part. Furthermore, this prescription must be such that, even though individual polylogarithms might develop an imaginary part, the individual basis functions $\cP_n$ and $\cQ_n$ remain real everywhere in the $(u,v)$ plane. In order to derive this prescription, let us start in region I and then continue the function to some other region. As long as we are in region I, $z$ and $\zp$ are complex conjugate to each other, and without loss of generality we may assume that
\beq
\textrm{Im}\,z>0 {\rm~~and~~} \textrm{Im}\,\zp<0\,.
\eeq
If we want the analytic continuation to be smooth, we have to choose the prescription
\begin{equation}\label{eq:prescription}
z \rightarrow z + i\varepsilon {\rm~~and~~} \zp \rightarrow \zp - i\varepsilon\,.
\end{equation}
Using this prescription we can work out all the analytic continuation formulae for the basis functions. The results are shown in appendix~\ref{app:functional_equations}. Note that none of the analytic continuation formulae shown in appendix~\ref{app:functional_equations} involves explicit factors of $i\pi$, which implies that all the basis functions are real everywhere in the $(u,v)$ plane, despite the fact that individual polylogarithms might develop an imaginary part. This is a direct consequence of the single-valuedness of the basis functions in region I.

Let us now concentrate on region II. In this region all the individual polylogarithms appearing inside the basis functions are real. Indeed, the basis functions have been chosen such that they are manifestly real if $0< z,\zp<1$. This condition is indeed satisfied, as we will show in the following.
It is easy to convince oneself that if one requires $u,v$ and $\lambda(1,u,v)$ to be positive, then there are only three possible ranges for $z$ and $\zp$,
\beq\label{eq:ranges}
z,\zp < 0\,, \qquad 0<z,\zp<1\,, \qquad z,\zp > 1\,.
\eeq 
Since $v<1$ in region II, we can see from the definition of $z$, eq.~\eqref{eq:z_definition}, that $z$ must be positive in region II, and $u<1$ then implies that we are in the range $0<z,\zp<1$.
As the basis functions are manifestly real for $0<z,\zp<1$, we conclude that the expressions we  derived for the master integrals in region I are still valid in region II.

Let us now turn to the regions III and IV. First, we note that these regions are related to region I
by a permutation of the external legs. In section~\ref{sec:SV_polylogs} we argued that permutations of the external legs act on the single-valued polylogarithms via the transformations of the argument~\eqref{eq:S3_symmetry}. For example, the mappings from region III or IV into region II are given by
\begin{eqnarray}
\textrm{III} \longleftrightarrow \textrm{II:} \quad  (u,v)\rightarrow (u/v,1/v) \quad& \Longleftrightarrow& \quad (z,\zp) \rightarrow (\zp/(\zp-1), z/(z-1))\,, \label{eq:III} \\
\textrm{IV} \longleftrightarrow \textrm{II:} \quad  (u,v)\rightarrow (1/u,v/u) \quad &\Longleftrightarrow& \quad (z,\zp) \rightarrow (1/\zp,1/z)\,.\label{eq:IV}
\end{eqnarray}
The first line corresponds to the interchange of the momenta $p_2$ and $p_3$, while the second line corresponds to the interchange of the momenta $p_1$ with $p_3$. We can therefore always map the results in the regions III and IV to the region II, where we know that all the polylogarithms are real.
This will be discussed in greater detail in the rest of this section.
We note in passing that it is of course also possible to directly map region III to IV (and vice versa), and   the integrals are even invariant under this transformation. This is expressed by interchanging $u$ with $v$, or equivalently $z \rightarrow 1-\zp$, which is just the permutation of $p_1$ with $p_2$. 

Let us now have a closer look at region III where $v>1$ and $z,\zp<0$. Indeed, if $\zp<0$, we have $1+u-v < \sqrt{\lambda(1,u,v)}$, and as $v$ is positive this condition is only fulfilled for $1+u-v<0$. From $v>u$ it then follows that we must be in region III. It is easy to see that the  transformation~\eqref{eq:III} maps $z,\zp<0$ to $0<\zp/(\zp-1),z/(z-1)<1$, i.e., region III is indeed mapped into region II. Note that the polylogarithms appearing inside the basis functions are also real for $z,\zp<0$, so that the expression valid in region I is still valid in region III. 

Let us finally turn to region IV. As this region is related to region II via the mapping $(z,\zp)\to(1-\zp,1-z)$, we immediately see that in region IV we have $z,\zp>1$. Individual polylogarithms appearing inside of the basis function may therefore develop imaginary parts, and we need to analytically continue them to region II via the transformation~\eqref{eq:IV}, using the prescription~\eqref{eq:prescription}.
The arguments of the functions $\cP_n$ and $\cQ_n$ can then be mapped into the unit interval by using the functional equations of appendix~\ref{app:functional_equations}. Let us illustrate this on the examples of the one-loop triangle $T_1$ and the two-loop master integral $F_1$. We can use the functional equations to rewrite the expressions for the integrals (\ref{eq:one_loop_triangle}, \ref{eq:I1}) such that $z$ appears as $1/z$ in the argument of the polylogarithms. We obtain

\begin{align}
\cI(1,1,1;u,v;\eps)_{|\textrm{Region IV}} &\,= 2{\,u^{-\eps}\,v^{-\eps}\over z-\zp}\,\Bigg\{-\cP_2(1/z) - 2\eps[\cQ_3(1/z)+\ln u\,\cP_2(1/z)] \\ \nonumber
& \,+ \eps^2\,\Big[\Big(\zeta_2 + \frac{1}{6} \log u \log v\Big)\cP_2(1/z)+2\, [-\cQ_4^-(1/z) - 2\ln u\,\cQ_3(1/z) \\ \nonumber &\,- \frac{13}{12}\ln^2 u\,\cP_2(1/z) + \frac{1}{6}\ln u\ln v\cP_2(1/z)]\Big]+\ord(\eps^3)\Bigg\}\,,
\end{align}
\beq\bsp
\cI&(\eps,1,1;u,v;\eps)_{|\textrm{Region IV}} = \frac{1}{1-3\eps}\,\Bigg\{\frac{1}{\eps} + \eps(1-u-v)\,\cI(1+\eps,1,1;u,v)_{|\textrm{Region IV}}\\
& - \eps\left[2\zeta_2 + \ln v\ln u\right] + \eps^2\left[-6\,\cP_3(1/z)-6\,\cP_3(1-z) + \frac{1}{2}\ln u\ln^2 v + \frac{1}{2}\ln^2 u \ln v + 6\,\zeta_3 \right] \\ 
&+ \eps^3\Big[\frac{3}{4}\,\cQ_4^+(1/z) + \frac{3}{4}\,\cQ_4^+(1-z) - \frac{9}{2}\,\zeta_2\ln^2 u + 3(3\ln u + \frac{1}{2}\ln v)\,\cP_3(1/z) \\ &\,+ \frac{3}{2}(\ln u - 3\ln v)\cP_3(1-z) -\frac{9}{2}\,\cP_2(1/z)^2 + \frac{9}{16}\ln^4 v - \frac{41}{8}\,\ln v\ln^3 u - \frac{25}{24}\,\ln u\,\ln^3 v \\ &\,+\frac{3}{8}\ln^2 u\ln^2 v -\,\zeta_2\,\ln u\ln v +3\,\zeta_3\,\ln u - 12\,\zeta_3\,\ln v + 7\zeta_4 \Big] \Bigg\} +\ord(\eps^4)\,.
\esp\eeq

\subsection{Analytic continuation to region V}
Let us conclude this section by discussing the results for the master integral in region V where $\lambda(1,u,v)=0$, or equivalently $z=\zp$.
While the master integrals are obviously finite in this region, 
they in general contain a prefactor
 $\frac{1}{z-\zp}$, which is obviously divergent in the limit $\zp\to z$. This pole must therefore be spurious, and in the following we discuss how we can take the limit $\zp\to z$ in order to obtain the analytic expression for the master integrals in region V.
 
 As an example, consider the two-loop master integral $F_3$ given in eq.~\eqref{eq:F3_result}. In order to obtain the analytic expression in region V we need to take a limit of the form
\beq\label{eq:hospital}
\lim_{\zp\to z}\frac{f(z,\zp) - f(\zp,z)}{z-\zp} = {\partial\over\partial\zp}f(z,\zp)_{|\zp=z}\,,
\eeq 
for some function $f(z,\zp)$. The numerator and denominator in the left-hand-side of eq.~\eqref{eq:hospital} vanish simultaneously in the limit, and we can apply L'Hospital's rule to evaluate the limit in terms of the derivative of $f(z,\zp)$.

As an example, applying eq.~\eqref{eq:hospital} to the analytic expression for $F_3$, eq.~\eqref{eq:F3_result}, we obtain,
\beq\bsp\label{eq:F3zzp}
F_3(\zp=z;\eps) = &\, c_\Gamma^2\,(-p_3^2)^{-1-2\eps}\,{6\over z(1-z)}\left[\li_3(z) - \li_2(z)\ln z - \frac{1}{3}\ln^2z\ln(1-z) - \frac{\zeta_3}{2}\right] \\ & + (z \leftrightarrow 1-z)+ \ord(\eps)\,.
\esp\eeq
We stress that eq.~\eqref{eq:F3zzp} is only valid for $z<1$. The corresponding results in the other regions can be obtained from the inversion relations of the previous section.
Furthermore, note that, unlike the expression for generic $(z,\zp)$, eq.~\eqref{eq:F3zzp} consists only of polylogarithms of transcendental weight three. This is an immediate consequence of the application of the differential operator in eq.~\eqref{eq:hospital}.
We checked eq.~\eqref{eq:F3zzp} numerically with \emph{Mathematica}.  
  
In the same way we can obtain the expressions of all other master integrals in region V.
In particular, all the terms that have a spurious at $z=\zp$ are all proportional to either
the one-loop triangle $T_1$ or the two-loop ladder triangle $F_3$.
It is therefore sufficient to compute the limit $\zp\to z$ of $T_1$ in addition to the result for $F_3$ we just obtained.
The computation of the limit~\eqref{eq:hospital} is, however, technically more involved in this case because of the
appearance of the new basis functions $\cQ_3(z)$ and $\cQ_4^-(z)$ which contain multiple polylogarithms.
Acting with the derivative on these multiple polylogarithms can generate new singular factors $\frac{1}{z-\zp}$, and we have to iterate 
eq.~\eqref{eq:hospital} until we reach a finite result. In the end, we arrive at
\beq\bsp\label{eq:T1_regV}
T_1(\zp=z;\eps) = &\,-2 c_\Gamma\,{\Gamma(1-2\eps)\over\Gamma(1-\eps)^2}\,(-p_3^2)^{-1-\eps} \frac{\ln z}{1-z} \Big[ -1 + \eps\big( \ln z - 2 \big) \\
&\,+ \eps^2\Big( -\frac{2}{3} \ln^2z + 2\ln z + \zeta_2 -4 \Big) \Big] + (z \leftrightarrow 1-z)+ \ord(\eps^3)\,.
\esp\eeq
Interestingly all the polylogarithms disappear in the limit and the result is expressed through logarithms only. We checked the correctness of eq.~\eqref{eq:T1_regV} by computing the integral directly from the Feynman-representation~\eqref{eq:I_definition} and letting $u=z^2$ and $v=(1-z)^2$ from the start. We find perfect agreement with the expression~\eqref{eq:T1_regV} obtained by differentiating the result for generic $(z,\zp)$. 

Having at our disposal the analytic results for $T_1$ and $F_3$ in region V, we can easily obtain the corresponding results for all other master integrals.
We start with the integral $F_4$. The result is given by

\beq
F_4(\zp=z;\eps) = 4\,c_{\Gamma}^2(-p_3^2)^{-2-2\eps}\frac{(z\ln z + (1-z)\ln(1-z))^2}{z^2(1-z)^2} + \ord(\eps).
\eeq
Next we turn to the integral $F_2$, which is a combination of $T_1$ and $F_3$,
\beq\bsp\
F_{2}(\zp=z;\eps) & = -\frac{c_\Gamma}{\eps}\,{\Gamma(1+2\eps)\over \Gamma(1+\eps)^2}\,(-p_3^2)^{-\eps}\,\left(\frac{z^{-2\eps}+(1-z)^{-2\eps}}{2} -\eps^2\,\zeta_2\right)\,T_1(\zp=z;\eps)\\
&\,+\eps\,{\Gamma(1+2\eps)\,\Gamma(1-2\eps)\over \Gamma(1+\eps)^2\,\Gamma(1-\eps)^2}\,F_{3}(\zp=z;\eps)+\ord(\eps^2)\,.
\esp\eeq
Finally, the integral $F_1$ is manifestly finite as $\zp\to z$, apart for the term proportional to $F_2$. We can therefore simply set $\zp=z$, and express all the multiple polylogarithms in terms of harmonic polylogarithms in $z$. We obtain
\beq\bsp
F_{1}(\zp=z;\eps) & = \frac{\eps}{2(1-2\eps)}\frac{z(1-z)}{1-3\eps}(-p_3^2)\,F_2(\zp=z;\eps) \\ &+ \frac{c_{\Gamma}^2}{2\eps(1-2\eps)} \frac{\Gamma(1+2\eps)\,\Gamma(1-2\eps)}{\Gamma(1+\eps)^2\,\Gamma(1-\eps)^2}(-p_3^2)^{-2\eps} \,\frac{1}{1-3\eps} \\ &\, \Bigg\{\frac{1}{2\eps} - \eps\left[\zeta_2 + 2\ln z\ln(1-z)\right] \\ &+ \eps^2\left[-12\li_3(z) + 12\li_2(z)\ln z + 8\,\ln^2 z\ln(1-z) + 3\,\zeta_3 \right] \\ &+ \eps^3\Bigg[-24\li_4(z) + 12\li_3(z)(3\ln z + \ln(1-z)) -24\li_2(z)\ln^2 z  \\ & \qquad -\frac{32}{3}\ln^3 z\ln(1-z) -12\zeta_3\ln z + \ln^2 z\ln^2(1-z)  \\ & \qquad -\frac{\pi^2}{3}\ln z\ln(1-z) + \frac{13}{2}\zeta_4\Bigg]\Bigg\}  + (z \leftrightarrow 1-z)+\ord(\eps^4)\,.
\esp\eeq
As for the one-loop integral, we checked our results for $F_1$ and $F_2$ by directly evaluating the Feynman parameter integral~\eqref{eq:I_definition} with $u=z^2$ and $v=(1-z)^2$.


\section{Conclusion}
In this paper we studied one and two-loop triangle integrals without internal masses and with three off-shell external legs.
Based on general arguments coming from symbols and the Hopf algebra of multiple polylogarithms, we have shown that 
in region I where the K\"allen function is negative the results can be expressed through certain single-valued multiple polylogarithms.
Our results in this region are characterized by a strikingly simple and compact analytic form. In particular, it turns out that the master integral $F_2$
can be written as a combination of the one-loop triangle and the two-loop master integral $F_3$, at least up to the order relevant to two-loop computations.
We have also studied the single-valued basis functions and we have shown that their functional equations almost trivialize the analytic continuation of our results
to all other regions where the external masses have the same sign. We have checked that all our results satisfy the differential equations of ref.~\cite{Birthwright:2004kk}. Furthermore, we checked that our expressions numerically\footnote{All multiple polylogarithms were evaluated numerically using GiNaC~\cite{Bauer:2000cp,Vollinga:2004sn}.} against {\tt FIESTA}~\cite{Smirnov:2008py,Smirnov:2009pb} for various points 
in the different regions. In addition, we have compared our results numerically against the two-equal-mass results of ref.~\cite{Gehrmann-Tancredi}. In all cases we found perfect agreement.

While this is not the first time that these master integrals have been considered in the literature, we believe that our results
go beyond previously known representations for these functions in several ways. First, while results for all master integrals expanded in $\eps$ up to terms of transcendental weight four had already been derived in ref.~\cite{Birthwright:2004kk},
these expressions involve complicated iterated integrals which are not manifestly related to multiple polylogarithms. Secondly, other representations available in the literature are either expressed through log-sine integrals, and thus only valid in specific kinematic regions (representations in other regions can be obtained, but require the analytic continuation of the log-sine integrals) or not expanded high enough in $\eps$. All our results
are expressed through multiple polylogarithms only and are expanded in $\eps$ up to transcendental weight four. In addition, the functional equations among the basis functions make the analytic continuation of our results to other regions very simple.

We conclude by mentioning that the approach we have used, i.e., the a priori determination the space of functions and their properties, is not restricted to single-valued
functions and three-mass triangle integrals. Once a set of entries that can appear inside the symbol of a given loop integral has been determined, we can in the same way determine 
the space of functions relevant to Feynman integrals with more scales. Studies in this direction for two-loop box integrals with two external masses are currently under investigation.

\section*{Acknowledgements}
The authors are grateful to T.~Gehrmann and L.~Tancredi for sharing their analytic results for the two-loop master integrals in the limit of two equal masses.
This work was supported by the ERC grant ``IterQCD''. All Feynman diagrams in this paper were drawn with the help of the axodraw package~\cite{axodraw}.

\appendix

\section{Construction of the basis functions}
\label{app:basis}
Our results for the three-mass triangles are expressed through the single-valued multiple polylogarithms briefly introduced in section~\ref{sec:SV_polylogs}. A basis for these functions is given by single-valued versions of HPLs, augmented by some new basis functions which involve $z-\zp$ as an entry in its symbol. While a generic construction of single-valued HPLs is known for arbitrary weight~\cite{BrownSVHPLs}, the new basis functions have, at least to our knowledge, never been studied in the literature. In this appendix we present a recursive procedure that allows to construct a basis for the space of single-valued functions whose symbol has entries drawn from the set $\cZ=\{z,\zp,1-z,1-\zp,z-\zp\}$.
While we only present the construction on the example of this particular space of functions, we emphasize that it is easy to extend the construction to more general classes of functions whose symbols have prescribed entries and satisfy certain first entry conditions. 

Let us denote by $\cH$ the space spanned by multiple zeta values and polylogarithmic functions whose symbols have entries drawn from the set $\cZ$, and by $\cH_{SV}$ the subspace of $\cH$ spanned by single-valued functions. As we will see below, the single-valuedness criterion is not entirely equivalent to the first entry condition: while single-valuedness implies the first entry condition, the inverse is not necessarily true. 

It is easy to check that $\cH_{SV}$ is in fact a subalgebra of $\cH$ (if we take the product of two functions with a prescribed set of discontinuities at most, the product will not have any new discontinuities). Furthermore, both $\cH$ and $\cH_{SV}$ are graded by the weight, i.e., they can be written as direct sums of vector spaces of polylogarithms of a given weight,
\beq
\cH=\bigoplus_{n=0}^\infty\cH_n {\rm~~and~~} \cH_{SV} = \bigoplus_{n=0}^\infty\cH_{SV,n}\,,
\eeq
where we put by convention $\cH_0=\cH_{SV,0}=\mathbb{Q}$. We furthermore decompose each graded subspace into eigenspaces of complex conjugation\footnote{We could of course decompose the space of functions further into eigenspace of the $S_3$ action~\eqref{eq:S3_symmetry}. In the following we prefer not to do this.} (i.e., we decompose every function into its real and imaginary part),
\beq
\cH_n = \cH_n^+\oplus\cH_n^-{\rm~~and~~} \cH_{SV,n} = \cH_{SV,n}^+\oplus\cH_{SV,n}^-\,.
\eeq
Our goal is to find, up to a given weight (four in our case), a basis for the different graded subspaces.

A first naive approach would consist in writing down the most general symbol of a given weight that satisfies the first entry condition and has the correct behavior under complex conjugation, and then to impose integrability of the symbol. Indeed, not every tensor corresponds to a function, but it can be shown that a tensor of the form
\beq
\sum_{I=(i_1,\dots,i_m)} c_I\,\omega_{i_1} \otimes \cdots \otimes \omega_{i_m}
\eeq
is the symbol of a function if and only if the following \emph{integrability condition} is fulfilled~\cite{symbolsC},
\beq\label{eq:integrability}
\sum_{I=(i_1,\dots,i_m)} c_I \  \big[d\log \omega_{i_j}\wedge d\log \omega_{i_{j+1}}  \big]\, \omega_{i_1} \otimes \cdots \otimes
 \widehat \omega_{i_j} \otimes  \widehat\omega_{i_{j+1}} \otimes
\cdots \otimes \ \omega_{i_m} \ = \ 0 \,,
\eeq
where the hat indicates that the corresponding entry in the symbol is omitted. This naive approach however leads to two problems:
\begin{enumerate}
\item For high weights, this approach is very inefficient, as it requires to impose the integrability condition at every weight for all the components. It is more efficient to `recycle' information from lower weight, as dropping the last entry of an integrable tensor with prescribed first entry produces a tensor with the same properties, but of lower weight.
\item This approach does not allow to incorporate multiple zeta values. However, as can be seen from the example of the single-valued HPLs of ref.~\cite{BrownSVHPLs}, individual terms in a single-valued combination might contain non-single-valued pieces proportional to multiple zeta values. These terms are thus missed in this approach, and the resulting function will not necessarily be an element of $\cH_{SV}$, despite the fact that its symbol satisfies the first entry condition. In other words, single-valuedness implies the first entry condition, but the inverse is not necessarily true. A fully symbol-based approach cannot incorporate this delicate distinction.
\end{enumerate}

We thus need a more refined way of constructing the basis functions. In ref.~\cite{Duhr:2012fh} it was shown that a way to incorporate zeta values into the symbol calculus is obtained by promoting the symbol map to the full coproduct of the Hopf algebra of multiple polylogarithms~\cite{Goncharov:2001}. 
In this setting the symbol is nothing but the maximal iteration of the coproduct. While $\cH$ is obviously a Hopf algebra, its single-valued subspace $\cH_{SV}$ is not, but we rather have\footnote{Technically speaking, $\cH_{SV}$ is an $\cH$-module.}
\beq\label{eq:comodule}
\cH_{SV}\stackrel{\Delta}{\longrightarrow} \cH_{SV}\otimes\cH\,.
\eeq
This follows immediately from the fact that the terms in the second factor do not need to satisfy the first entry condition, or more generally, from the fact that only the first factor in the coproduct embodies the information about the discontinuities of the functions~\cite{Duhr:2012fh}.

Eq.~\eqref{eq:comodule} allows us to formulate a recursive procedure to construct a basis for $\cH_{SV}$. The basis at weight one is obviously known, both for $\cH_1$ and it single-valued subspace $\cH_{SV,1}$,
\beq\bsp
\cB_1^+ = \{\ln|1-z|^2,\ln|z|^2,\ln(z-\zp)\}\,,&\qquad\cB_1^-=\{\ln\frac{1-z}{1-\zp},\ln\frac{z}{\zp}\}\,,\\
\cB_{SV,1}^+= \{\ln|1-z|^2,\ln|z|^2\}\,,&\qquad\cB_{SV,1}^-=\emptyset\,.
\esp\eeq
Next, suppose that we know a basis  of $\cH_{SV,n}^\pm$. We will show in the following how we can use this basis and recursively construct from eq.~\eqref{eq:comodule} a basis for $\cH_{SV,n+1}^\pm$. The idea is simple: rather than constructing a basis for $\cH_{SV,n+1}^\pm$, we construct a basis of $\cH_{SV,n}\otimes \cH_1$ instead and then `lift' this basis to a basis of $\cH_{SV,n+1}^\pm$ using the integrability condition. Note however that in this way we can only obtain basis elements of $\cH_{SV,n+1}^\pm$ which have a non-zero image in $\cH_{SV,n}\otimes \cH_1$ under the coproduct. It is easy to see that those elements of $\cH_{SV,n+1}^\pm$ with vanishing image in $\cH_{SV,n}\otimes \cH_1$ are precisely the \emph{primitive elements} in $\cH_{SV}$, i.e., elements $\xi\in\cH_{SV}$ such that 
\beq
\Delta(\xi) = 1\otimes\xi + \xi\otimes 1\,.
\eeq
We know that in the present case the primitive elements of weight $n>1$ that are of interest are precisely the zeta values of depth one, $\zeta_n$. Hence, our construction will produce a set of basis element to which we need to add at each weight the corresponding zeta value of depth one in order to get a full basis. 

Let us now assume that we are given a basis $\cB_{SV,n}^\pm\equiv\{b_{n,i}^\pm\}_{1\le i\le d_n^\pm}$ of $\cH_{SV,n}^\pm$ ($d_n=\textrm{dim}_\mathbb{Q}\cH_{SV,n}^\pm$). Writing $\Delta_{n,1}$ for the component of the coproduct that takes values in $\cH_n\otimes\cH_1$, and using the fact that
\beq\bsp\label{eq:decomposition}
\cH_{SV,n+1}^+&\,\stackrel{\Delta_{n,1}}{\longrightarrow} (\cH_{SV,n}^+\otimes\cH_1^+) \oplus (\cH_{SV,n}^-\otimes\cH_1^-)\,,\\
\cH_{SV,n+1}^-&\,\stackrel{\Delta_{n,1}}{\longrightarrow} (\cH_{SV,n}^+\otimes\cH_1^-) \oplus (\cH_{SV,n}^-\otimes\cH_1^+)\,,
\esp\eeq
we see that $\cB_n^{++}\cup \cB_n^{--}$ and 
$\cB_n^{+-}\cup \cB_n^{-+}$, with 
\beq
\cB_n^{s_1,s_2} = \left\{b_1\otimes b_2: b_1\in\cB_{SV,n}^{s_1}\textrm{ and } b_2\in\cB_1^{s_2}\right\}\,,
\eeq
form a basis for the spaces in the right-hand side of eq.~\eqref{eq:decomposition}. However, not every basis element lies in the image of $\Delta_{n,1}$. We thus need to construct the correct linear combinations. This is achieved by writing down the most general linear combination of vectors in each $\cB_n^{s_1,s_2}$ and acting with $\Delta_{n-1,1}\otimes\textrm{id}$
and imposing the integrability condition~\eqref{eq:integrability} in the last two factors. Formally, if we consider for example a generic linear combination $\xi$ of the basis elements in $\cB_n^{++}\cup \cB_n^{--}$, then the requirement for $\xi$ to lie in the image of $\Delta_{n,1}$, i.e., for $\xi$ to be integrable, reads,
\beq\label{eq:integ}
(\textrm{id}\otimes d\wedge d)(\Delta_{n-1,1}\otimes\textrm{id})(\xi) = 0\,,
\eeq
where $d$ denotes the usual differential on differential forms. Eq.~\eqref{eq:integ} gives rise to a linear system for the coefficients in the linear combination $\xi$, and the solution space of this equation is related to the image of $\cH_{SV,n+1}^\pm$ under $\Delta_{n,1}$. The next step is simply to find functions whose image under $\Delta_{n,1}$ matches the solutions to eq.~\eqref{eq:integ}, a step which in this case can easily be carried out using the methods of ref.~\cite{Duhr:2011zq}. 

The functions obtained in this way are not necessarily single-valued, despite the fact that their symbols satisfy the first entry condition. The reason for this lies in the fact the we lost track of all terms proportional to $\zeta_n$ when acting with $\Delta_{n-1,1}$. In the following we describe how we can uniquely fix this ambiguity in the present case. Indeed, as zeta values are real, we only have to deal with two different cases:
\begin{enumerate}
\item In the odd sector, $\cH_{SV,n+1}^-$, the terms we lost in eq.~\eqref{eq:integ} are necessarily of the form
\beq\label{eq:odd_ansatz}
\alpha\,\zeta_n\,\ln\frac{z}{\zp} + \beta\,\zeta_n\,\ln\frac{1-z}{1-\zp}\,,\qquad \alpha\,\beta\in\mathbb{Q}\,.
\eeq
While these functions are not single-valued by themselves, they can appear together with other polylogarithms in a single-valued combination (see for example the last term in eq.~\eqref{eq:Q3_def}).
It turns out that for each independent solution to eq.~\eqref{eq:integ} we can fix $\alpha$ and $\beta$ in a unique way. Indeed, all the functions in $\cH_{SV,n+1}^-$ have the property that they must vanish when $z$ approaches the real axis, and so they must in particular vanish if both $z$ and $\zp$ approach the points 0 or 1. The solutions to eq.~\eqref{eq:integ} contain terms that diverge in this limit, and $\alpha$ and $\beta$ are then fixed by requiring the divergence to cancel.
\item In the even sector, $\cH_{SV,n+1}^+$, the relevant terms are
\beq
\alpha\,\zeta_n\,\ln|z|^2 + \beta\,\zeta_n\,\ln|1-z|^2 + \gamma\,\zeta_n\,\ln(z-\zp)\,, \qquad \alpha\,\beta\,\gamma\in\mathbb{Q}\,.
\eeq
The first two functions are manifestly single-valued by themselves, so we do not need to fix the coefficients $\alpha$ and $\beta$, but we simply add these functions to our basis. The remaining coefficient $\gamma$ is fixed in a similar way as in the odd sector, by requiring the function to have a smooth limit as $z$ becomes real (note that the function is no longer required to vanish in this case).
\end{enumerate}

Following this procedure, we can in principle construct a basis for $\cH_{SV,n}^\pm$ for every weight $n$. In practice, the combinatorics increases very quickly, which renders the procedure quickly inefficient. The main reason for this combinatorial problem lies in the fact that the number of basis elements increases very quickly with $n$. The construction can however easily be carried out up to weight four, and as a result we obtain all the functions shown in tab.~\ref{tab:indecomposables} (plus all possible products among these functions). The first time a single-valued function appears that cannot be expressed through single-valued HPLs alone is at weight three, where the new function $\cQ_3(z)$ appears. This function was already defined in eq.~\eqref{eq:Q3_def}. At weight four, we find three new functions, two of which are simply related by replacing $z$ by $1-z$. The two new independent functions read
\beq\bsp
\cQ&_4^+(z)\, =-4 \left[G\left(0,\frac{1}{z},0,\frac{1}{\zp},1\right)+G\left(0,\frac{1}{\zp},0,\frac{1}{z},1\right)\right]-4 \left[G\left(0,\frac{1}{z},0,\frac{1}{z},1\right)+G\left(0,\frac{1}{\zp},0,\frac{1}{\zp},1\right)\right]\\
&-4 \log|z|^2 \left[G\left(0,\frac{1}{z},\frac{1}{\zp},1\right)+G\left(0,\frac{1}{\zp},\frac{1}{z},1\right)\right]+[\text{Li}_2(z)-\text{Li}_2(\zp)]^2+2 [\text{Li}_2(z)+\text{Li}_2(\zp)]^2\\
&-16 [\text{Li}_4(z)+\text{Li}_4(\zp)]-7 [\text{Li}_2(z)+\text{Li}_2(\zp)]\, \log^2|z|^2+3 [\text{Li}_2(z)-\text{Li}_2(\zp)] \,\log\frac{1-z}{1-\zp} \log|z|^2\\
&+3 [\text{Li}_2(z)+\text{Li}_2(\zp)]\,\log|1-z|^2\, \log|z|^2+6 [\text{Li}_3(1-z)+\text{Li}_3(1-\zp)]\, \log|z|^2\\
&+18 [\text{Li}_3(z)+\text{Li}_3(\zp)]\, \log|z|^2-\frac{3}{4} \log^4|z|^2+\frac{3}{4} \log^2 \frac{1-z}{1-\zp} \log^2|z|^2+\frac{3}{4} \log^2|1-z|^2\, \log^2|z|^2\\
&\,+\frac{3}{2} \log\frac{1-z}{1-\zp} \,\log|1-z|^2\, \log \frac{z}{\zp}\, \log|z|^2\,.
\esp\eeq
\beq\bsp
\cQ_4^-(z)&\, =-G\left(0,0,\frac{1}{z},\frac{1}{\zp},1\right)+G\left(0,0,\frac{1}{\zp},\frac{1}{z},1\right)+\frac{1}{2} \left[G\left(0,\frac{1}{z},0,\frac{1}{z},1\right)-G\left(0,\frac{1}{\zp},0,\frac{1}{\zp},1\right)\right]\\
&+\frac{1}{2} \left[G\left(0,\frac{1}{\zp},\frac{1}{z},\frac{1}{z},1\right)-G\left(0,\frac{1}{z},\frac{1}{\zp},\frac{1}{\zp},1\right)\right]+\frac{1}{2} \left[G\left(0,\frac{1}{\zp},\frac{1}{z},\frac{1}{\zp},1\right)-G\left(0,\frac{1}{z},\frac{1}{\zp},\frac{1}{z},1\right)\right]\\
&+\frac{1}{2} \left[G\left(0,\frac{1}{\zp},\frac{1}{\zp},\frac{1}{z},1\right)-G\left(0,\frac{1}{z},\frac{1}{z},\frac{1}{\zp},1\right)\right]+\frac{1}{8} \log ^2|z|^2 \left[G\left(\frac{1}{z},\frac{1}{\zp},1\right)-G\left(\frac{1}{\zp},\frac{1}{z},1\right)\right]\\
&+\frac{1}{4} \log|z|^2\,\log|1-z|^2\, \left[G\left(\frac{1}{z},\frac{1}{\zp},1\right)-G\left(\frac{1}{\zp},\frac{1}{z},1\right)\right]-\frac{1}{4} [\text{Li}_2(z)-\text{Li}_2(\zp)] [\text{Li}_2(z)+\text{Li}_2(\zp)]\\
&-\frac{1}{2} \log|z|^2 \left[G\left(\frac{1}{z},\frac{1}{z},\frac{1}{\zp},1\right)-G\left(\frac{1}{\zp},\frac{1}{\zp},\frac{1}{z},1\right)\right]+\frac{1}{2} [\text{Li}_4(1-\zp)-\text{Li}_4(1-z)]\\
&+2 [\text{Li}_4(z)-\text{Li}_4(\zp)]+\frac{1}{16} [\text{Li}_2(z)-\text{Li}_2(\zp)]\,\log^2\frac{1-z}{1-\zp}+\frac{1}{48} \log\frac{z}{\zp}\, \log^3|1-z|^2\\
&-\frac{1}{12} [\text{Li}_2(z)-\text{Li}_2(\zp)]\, \log|z|^2\, \log|1-z|^2+\frac{1}{8} [\text{Li}_2(z)+\text{Li}_2(\zp)]\,\log\frac{1-z}{1-\zp}\,\log|1-z|^2\\
&+\frac{1}{4} [\text{Li}_3(1-z)-\text{Li}_3(1-\zp)] \,\log|1-z|^2+\frac{1}{16} [\text{Li}_2(z)-\text{Li}_2(\zp)]\, \log^2|1-z|^2\\
&+\frac{1}{4} [\text{Li}_3(1-z)+\text{Li}_3(1-\zp)] \,\log\frac{1-z}{1-\zp}+\frac{1}{24}\log\frac{1-z}{1-\zp} \,\log ^3|z|^2\\
&+\frac{1}{48} \log\frac{1-z}{1-\zp}\, \log ^2|z|^2 \,\log|1-z|^2+\frac{1}{8}\log\frac{1-z}{1-\zp}\, \log|z|^2\, \log^2|1-z|^2\\
&+\frac{1}{16} \log^2\frac{1-z}{1-\zp}\,  \log\frac{z}{\zp}\, \log|1-z|^2-\frac{1}{4} \zeta_2\, \log\frac{1-z}{1-\zp}\, \log |1-z|^2\,.
\esp\eeq 

Let us conclude this section by discussing whether or not the functions $\cQ^\pm_4(z)$ we just introduced can be expressed through classical polylogarithms alone. If we define $\delta$ as the linear operator acting on tensors of rank four via
\beq
\delta(a\otimes b\otimes c\otimes d) = (a\wedge b)\wedge(c\wedge d)\,,
\eeq
with $a\wedge b = a\otimes b-b\otimes a$, then it follows from the conjecture of ref.~\cite{Goncharov_delta} that a combination $f$ of multiple polylogarithms of weight four can be expressed through classical polylogarithms only if and only if the symbol of $f$ satisfies $\delta\left[\cS(f)\right] = 0$. Acting with $\delta$ on $\cQ^\pm_4(z)$, we find
\beq
\delta\left[\cS\left(\cQ^+_4(z)\right)\right]\neq0 {\rm~~and~~}\delta\left[\cS\left(\cQ^-_4(z)\right)\right]=0\,.
\eeq
We thus conclude that we cannot express $\cQ^+_4(z)$ through classical polylogarithms alone. $\cQ^-_4(z)$ on the other hand could be expressed through classical polylogarithms only. The result would consist in a very complicated combination of classical polylogarithms of weight four, each of which has very  complicated branch cuts which cancel mutually. Just like in the case of $\cQ_3^-(z)$, we refrain for this reason from giving the expression of $\cQ^-_4(z)$ in terms of classical polylogarithms.

\section{Functional equations for the basis functions}
\label{app:functional_equations}
In this section we summarize the functional equations satisfied by the single-valued functions shown in tab.~\ref{tab:indecomposables}. The functional equations we consider are those which arise through the action of the symmetric group $S_3$ defined by eq.~\eqref{eq:S3_symmetry}. The group $S_3$ is generated by two elements, which we choose to correspond to 
\beq\label{eq:gens}
z\to 1/\zp {\rm~~and~~} z\to 1-\zp\,.
\eeq
We therefore only show functional equations for these two transformations, and all others can be obtained by iterating the functional equations corresponding to transformations in eq.~\eqref{eq:gens}. The relevant functional equations for the single-valued analogues of the classical polylogarithms have already been given in eqs.~\eqref{eq:Pn_inversion} and~\eqref{eq:P2_reflection}. The functional equations for the new basis functions are
\beq\bsp
\cQ_3(1-\zp) &\,= \cQ_3(z)\,,\\
\cQ_3(1/\zp) &\,= \cQ_3(z)-\ln|z|^2\,\cP_2(z)\,,
\esp\eeq

\beq\bsp
\cQ_4^+(1/\zp) &\,= \cQ_4^+(z)-\frac{3}{2} \log|1-z|^2\, \log ^3|z|^2-12\, \zeta_2\, \log|1-z|^2\, \log |z|^2\\
&\,-18 \, \log |z|^2\,\cP_3(z)- 20\, \zeta_3\, \log|z|^2 + \frac{3}{4}\, \log ^4|z|^2+6\,\zeta_2\, \log ^2|z|^2\,,\\
\cQ_4^-(1-\zp) &\,= \cQ_4^-(z)\,,\\
\cQ_4^-(1/\zp) &\,= \cQ_4^-(z)-2 \log|z|^2\, \cQ_3(z)+\frac{11}{12}\,\log ^2|z|^2\, \cP_2(z)\\
&\,+\frac{1}{6} \,\log|z|^2\,\log|1-z|^2\, \cP_2(z)\,.
\esp\eeq

\end{document}